

\input amstex
\documentstyle{amsppt}
\redefine\frak{\bold}
\redefine\ltimes{\tilde\times}

\magnification=1200
\parskip 12pt
\pagewidth{5.4in}
\NoRunningHeads
\expandafter\redefine\csname logo\string@\endcsname{}
\NoBlackBoxes

\redefine\ll{\lq\lq}
\redefine\rr{\rq\rq\ }
\define\rrr{\rq\rq}

\define\nat{{}^\natural}

\redefine\Bbb{\bold}
\define\C{\Bbb C}
\define\Cn{\C^n}
\define\R{\Bbb R}
\define\Z{\Bbb Z}

\define\CPn{\C P^n}
\define\CP{\C P}
\define\CS{\C^\ast}

 \define\al{\alpha}
\define\be{\beta}
\define\de{\delta}
\define\la{\lambda}
\define\ep{\varepsilon}
\define\ga{\gamma}

\define\La{\Lambda}
\define\Lap{\Lambda_+}
\define\Lam{\Lambda_-}

\define\glnc{GL_n(\C)}
\define\grkcn{Gr_k(\C^n)}

\define\gc{G^{\C}}
\define\lieg{\frak g}
\define\liegc{\frak g^{\C}}
\define\liegp{\lieg_P}

\define\liekc{\frak k^{\C}}

\define\liehc{\frak h^{\C}}

\define\fl{\Lambda}

\define\blg{\Omega G}
\define\ablg{\Omega_{\alg}G}
\define\flg{\Lambda G}
\define\flgc{\Lambda G^{\C}}
\define\flpgc{\Lambda_+ G^{\C}}

\define\blu{\Omega U_n}

\define\lglnc{\La GL_n(\C)}
\define\ablu{\Omega_{\alg}U_n}
\define\alu{\La_{\alg}U_n}
\define\alglnc{\La_{\alg}GL_n(\C)}

\define\flgg{\Lambda\frak g}

\redefine\deg{\operatorname {deg}}
\redefine\det{\operatorname {det}}
\redefine\Im{\operatorname {Im}}
\redefine\Ker{\operatorname {Ker}}
\define\rank{\operatorname {rank}}
\redefine\dim{\operatorname {dim}}

\define\alg{\operatorname {alg}}
\define\Map{\operatorname {Map}}
\define\Hol{\operatorname {Hol}}
\redefine\Span{\operatorname {Span}}
\redefine\exp{\operatorname{exp}}
\define\Ad{\operatorname{Ad}}
\define\ad{\operatorname{ad}}
\define\HH{\operatorname{HH}}
\define\Hom{\operatorname {Hom}}

\define\ddz{\frac{\partial}{\partial z}}
\define\ddzb{\frac{\partial}{\partial\bar z}}
\redefine\sec{C^\infty}
\define\lra{\longrightarrow}
\define\gr{Gr_{kn-k}(\C^{kn})}
\redefine\sub{\subseteq}

\topmatter
\title
Group actions and deformations for harmonic maps
\endtitle
\author
Martin A. Guest and Yoshihiro Ohnita
\endauthor
\endtopmatter

 \document

\heading Introduction
\endheading

{}From the theory of integrable systems it is known that harmonic
maps from a Riemann surface to a Lie group may be studied by
infinite dimensional methods (cf. ~\cite{ZM},\cite{ZS}).
This was clarified
considerably by the papers \cite{Uh},\cite{Se}, especially in the case of
maps
from the Riemann sphere $S^2$ to a unitary group $U_n$.
The basic connection with infinite dimensional
methods is the correspondence between harmonic maps
$S^2\longrightarrow G$
and \ll extended solutions\rr $S^2\longrightarrow\blg$, where $G$ is any
compact Lie group and $\blg$ is its (based) loop group. In \cite{Uh} this
was used in two ways (in the case $G=U_n$):

\noindent (1) to introduce a {\it group action} of matrix valued rational
functions on
harmonic maps, and

\noindent (2) to prove a {\it factorization theorem} for harmonic
maps, which unifies and extends many of the known results on the
classification of harmonic maps from $S^2$ into various
homogeneous spaces.

\noindent In \cite{Se} it was shown that the factorization theorem can be
proved very naturally by using the \ll Grassmannian model\rr of
$\blg$, which is an identification of $\blg$ with a certain infinite
dimensional
Grassmannian (see \cite{PS}). In this paper we shall show how the group
action may be interpreted in terms of the Grassmannian model. The
advantages of this point of view are that the geometrical nature of
the action is emphasized, and that calculations become
easier. We shall illustrate this by giving some applications to
deformations of harmonic maps. By using some elementary ideas
from Morse theory, we obtain new results on the connectedness of
spaces of harmonic maps, a subject which has been studied recently
by various {\it ad hoc} methods (for example,
in \cite{Ve1},\cite{Ve2},\cite{Ve3},\cite{Lo},\cite{Kt}).

The paper is arranged as follows. In \S 1 we give the basic
definitions, including that of a \ll generalized Birkhoff
pseudo-action\rrr. The latter is an action of $k$-tuples of loops
$\ga$ on extended solutions $\Phi$, denoted by
$(\ga,\Phi)\longmapsto\ga^\sharp\Phi$. This definition involves a
Riemann-Hilbert factorization (a generalization of the Birkhoff
factorization for loops), and is an example of a \ll dressing
action\rr in the theory of integrable systems. Because the
factorization cannot always be carried out, the action is defined
only for certain $\ga$ and $\Phi$, so we call it a pseudo-action.
Nevertheless, it is possible to establish some general properties of
the action by using contour integral formulae, and we shall use these to
show that the most important case of a generalized Birkhoff
pseudo-action is precisely the one introduced by Uhlenbeck. In \S 2
we go on to show that the Uhlenbeck action on harmonic maps
$S^2\longrightarrow U_n$ of fixed energy \ll collapses\rr to the
pseudo-action of a finite dimensional group.
This collapsing phenomenon has been described from a different point
of view in \cite{AJS},\cite{AS1},\cite{AS2},\cite{JK}.

The Grassmannian model and its relevance for harmonic maps are
reviewed in \S 3. From this point of view there is a natural action of
the complex group $\flgc$ on extended solutions, where $\gc$ is the
complexification of $G$ and $\flgc$ is its (free) loop group. This
action is denoted by $(\ga,\Phi)\longmapsto\ga\nat\Phi$.  Elementary
properties of this action -- which really is an action, not a
pseudo-action -- are given in \S 4. In particular, it is easy to see
that this action, like the Uhlenbeck action, collapses to an action of
a finite dimensional Lie group.

Our first main result appears in \S 5, where we show that the actions
$\sharp$ and $\natural$ are
essentially the same, despite their very different definitions. The
essential point here is that the explicit Riemann-Hilbert
factorization needed for $\sharp$ is incorporated into the definition
of the Grassmannian needed for $\natural$. This result explains the
similarities between the properties of the action $\sharp$
(described in \S1 and \S 2) and the properties of the action
$\natural$ (described in \S 3 and \S 4). In particular, it \lq\lq
explains\rq\rq\  and extends Theorem 9.4 of [Uh].

In \S 6, we discuss applications of the action $\natural$ to
deformations of harmonic maps. A one parameter subgroup
$\{\ga_t\}$ of $\flgc$ gives rise to a deformation
$\Phi_t=\ga_t\nat\Phi$ of an extended solution $\Phi$. This
deformation has a simple geometrical interpretation: it is the result
of applying the gradient flow of a suitable Morse-Bott function on
$\blg$ to the extended solution $\Phi$. Hence, we obtain a new
extended solution $\Phi_\infty=\lim_{t\to\infty}\Phi_t$ which takes
values (almost everywhere) in a critical manifold of this Morse-Bott
function. In general,
$\Phi_\infty$ has a finite number of (removable) singularities. This
illustrates the well known fact (see \cite{SU}) that a sequence of
harmonic maps (of $S^2$) has a convergent subsequence over the
complement of a finite set, the latter being points at
which \ll bubbling off\rr occurs. We shall give some examples where
the singularities do not occur, so that $\Phi_\infty$ is joined to
$\Phi$  by a continuous path in the space of extended solutions.  The main
example is the following. Let  $\varphi:S^2\longrightarrow U_n$ be a
harmonic map, with corresponding normalized extended solution
$\Phi=\sum_{\al=0}^{m}T_{\al}\la^{\al}$ (this notation will be explained
later). Then we have (see Theorem 6.2):

\noindent (A) Assume that $\rank\,T_0(z)\ge2$ for all $z$. Then $\varphi$
can be deformed continuously to a harmonic map $\psi:S^2\longrightarrow
U_{n-1}$.

It is well known that harmonic maps into an inner symmetric space $G/K$
may be studied as a special case of harmonic maps into $G$ (by making use
of a totally geodesic embedding of $G/K$ into $G$). So our method can be
used to produce continuous deformations of harmonic maps from $S^2$ to
$G/K$, for various $G/K$. We shall give two examples, namely $G/K=\CP^n$
and $G/K=S^n$. In the first case we shall show:

\noindent (B) The number of connected components of the space of harmonic
maps $S^2\longrightarrow\CP^n$ is independent of $n$, if $n\ge2$.

\noindent This can be obtained as a consequence of the method for (A), but
we shall also give a direct proof (Theorem 6.5).
We conjecture that the space of harmonic maps $S^2\to\CP^n$ of fixed
energy and degree is connected. By (B), it would suffice to verify this
conjecture in the case $n=2$. In the case $G/K=S^n$, for $n\ge4$, we shall
use the same method to give a new proof of the following fact (Theorem 6.7;
see also \cite{Lo},\cite{Ve3},\cite{Kt}):

\noindent (C) The space of harmonic maps $S^2\longrightarrow
S^n$ of fixed energy is connected.

\noindent The proof we give is quite elementary and does not depend on
\S1-\S5 of this paper (though it was motivated by the method used for (A)).

Most of our results in \S 6 generalize to the case of extended solutions
$M\longrightarrow \blg$, where $M$ is any compact connected Riemann
surface. In particular, the results on the connected components of harmonic
maps from $S^2$ into $S^n$ or $\CPn$ generalize to the case of
{\it isotropic} harmonic maps into $S^n$ or {\it complex isotropic} harmonic
maps into $\CPn$. In fact, since our
method primarily involves the target space, one may go even further
and obtain similar results on pluriharmonic maps of compact connected
complex manifolds (cf. \cite{OV}).

Finally, we make some concluding remarks on the two main ingredients of
this paper, i.e. group actions and deformations. First, it should be
emphasized that the group actions discussed here do not represent a new
idea. It is a well known principle in other contexts to convert from real to
complex geometry, in order to reveal a larger (complex) symmetry group.
(Here, one converts from harmonic maps into a Riemannian manifold to
\lq\lq horizontal\rq\rq\  holomorphic maps into a complex manifold.)
Indeed, as mentioned above, the action $\sharp$ had its origins in the theory
of integrable systems, while examples of the action $\natural$ have been
treated explicitly in \cite{Gu} and have been alluded to by other authors. Our
contribution to this topic (in \S5) is the unification of the two actions.
Second, the results of \S6 concerning deformations are essentially
independent of \S1-\S5, although we feel that the group action provides
some motivation for these deformations. From a practical point of view, the
deformations have two main features. One is the connection with Morse
theory which allows us to predict easily the end result of the deformations.
The other is that the horizontality condition, which is sometimes hard to
deal with directly, is never needed explicitly in our calculations.

{\it Acknowledgements:} Our results on the connectedness of spaces of
harmonic maps were inspired by work of N. Ejiri and M. Kotani (cf. \cite{EK}).
The first author is indebted to W. Richter for pointing out the importance of
doing Morse theory on finite dimensional subvarieties of the loop group
(cf. \cite{Ri}). He acknowledges financial support from the Japan
Society for Promotion of Science and the U.S. National Science
Foundation.

$${}$$

\heading
\S{1}. Extended solutions and generalized Birkhoff pseudo-actions
\endheading
Let $M$ be a connected Riemann surface or, more generally, a
connected complex manifold. Let $G$ be a
compact connected Lie group equipped with a bi-invariant Riemannian
metric and let ${\frak g}$ denote its Lie algebra.
If necessary, we
choose a realization for the complexification  $\gc$ of $G$ as a
subgroup of some general linear group $\glnc$, with
$G=G^{\bold C}\cap{U(n)}$.
Let
$\mu$
denote the Maurer-Cartan form of
$G^{\bold C}$.
For a smooth map
$\varphi : M\longrightarrow G^{\bold C}$,
set
$\varphi^{\ast}\mu
=\al=\al^{\prime}+\al^{\prime\prime}$,
where $\al^{\prime}$ and $\al^{\prime\prime}$ are the
$(1,0)$-component and $(0,1)$-component of $\al$, respectively.

\noindent{\it Definition.} The map $\varphi:M\longrightarrow G^{\C}$ is
said to be {\it (pluri)harmonic} if and only if
$\bar{\partial}\al^{\prime}=\partial\al^{\prime\prime}.$

\noindent If $\varphi(M)\subseteq G$, then this definition coincides with
the usual definition (see (8.5) of \cite{EL}, \S 2 of \cite{OV}). We shall call
such a map $\varphi$ a {\it real harmonic map}.

For each
$\la\in{\C}^{\ast}={\C}\setminus\{0\}$,
consider the $1$-form on $M$ with values in ${\frak g}^{\C}$ given by
$$
\al_{\la}=
{1 \over 2}(1-\la^{-1})\al^{\prime}+
{1 \over 2}(1-\la)\al^{\prime\prime},
$$
and consider the first order linear partial differential equation
$$
{\Phi_{\la}}^{\ast}\mu=\al_{\la},
\tag{$\ast$}
$$
for a map $\Phi_{\la} : M\longrightarrow \gc$. Using an embedding
$\gc\longrightarrow\glnc$, this equation may be written as
$$
\cases
\partial{\Phi}_{\la}=
{1 \over 2}(1-\la^{-1})\Phi_{\la}\al^{\prime}\\
\bar\partial{\Phi}_{\la}=
{1 \over 2}(1-\la)\Phi_{\la}\al^{\prime\prime}.
\endcases
\tag{$\ast\ast$}
$$

\noindent{\it Definition.} A family of solutions $\Phi_{\la},
\la\in{\C}^{\ast}$, to ($\ast$) or ($\ast\ast$) is called an {\it extended
solution} (\cite{Uh}) or an {\it extended (pluri)harmonic map} (\cite{OV}).

\noindent The fundamental observation, proved in [Uh] for harmonic maps,
and extended
in [OV] to pluriharmonic maps, is:

\proclaim{Theorem 1.1}
Assume that $\Hom(\pi_1(M),G)=\{e\}$.
Choose a base point $z_0$ of $M$ and a map $\sigma :
{\C}^{\ast}\longrightarrow \gc$. Let $\varphi : M\longrightarrow G^{\C}$
be a (pluri)harmonic map. Then there exists a unique extended solution
$\Phi : M\times{\C}^{\ast}\longrightarrow \gc$
such that $\Phi_{\la}(z_0)=\sigma(\la)$.
Conversely, if $\Phi$ is an extended solution, then
$\Phi_{-1} :
 M\longrightarrow G^{\bold C}$
is a (pluri)harmonic map.
\qed
\endproclaim

\noindent Moreover, the extended solution $\Phi$ (obtained from $\sigma$
and $\varphi$) necessarily satisfies
$\Phi_{-1}=a{\varphi}$, where $a=\sigma(-1)\varphi(z_0)^{-1}$.

Let $\varphi$ be a real harmonic map.
If we choose $\sigma$ satisfying $\sigma(1)=e$ and
$\sigma(S^1)\subseteq{G}$, then $\Phi_1\equiv e$ and
$\Phi_{\la}(M)\subseteq G$ for any
$\la\in S^1=\{\la\in{\C}^{\ast}{\ }|{\ }|\la|=1\}$. (For example, we may
choose $\sigma\equiv e$.) In this case we call $\Phi$ a {\it real extended
solution}.

The smooth loop group of $G$ is defined by:
$$
\Omega{G}=\{\ga : S^1\longrightarrow G\ \vert\  \ga \text{ smooth},
\ga(1)=e\}.
$$
Let $\pi:\blg\longrightarrow G$ be the map $\pi(\ga)=\ga(-1)$.
A real extended solution $\Phi$ can be considered as a map into
$\Omega{G}$; conversely, if $\Phi:M\longrightarrow\blg$ satisfies ($\ast$)
or ($\ast\ast$) for $\la\in S^1$, then the same argument as for Theorem 1.1
shows that the map
$\varphi=\pi\circ\Phi : M\longrightarrow G$ is (pluri)harmonic. Because of
this we shall (with abuse of notation) use the term \ll real extended
solution\rr for any map $\Phi:M\longrightarrow\blg$ satisfying ($\ast$) or
($\ast\ast$).

It is known that $\Omega{G}$ has the structure of an infinite dimensional
homogeneous K\"ahler manifold (see \cite{PS}). There is a left-invariant
complex structure $J$ such that the $(+i)$-eigenspace of $J$ is the
subspace spanned by the elements
$(\la^{-k}-1){\frak g}^{\C}{\ }(k=1,2,\dots)$, under the
identification $T^{\C}_e\Omega{G}\cong\Omega\liegc$.
The condition ($\ast$) or ($\ast\ast$) may be written
$$
\Phi^\ast\mu(T_{1,0}M)
=\Phi^{-1}d\Phi (T_{1,0}M)
\subseteq
(\la^{-1}-1){\frak g}^{\C}.
$$
In particular, we see that any extended solution
$\Phi : M\longrightarrow \Omega{G}$ is holomorphic relative to $J$.

Following \cite{Uh}, we say that a harmonic map $\varphi$ has finite uniton
number  if there is an extended solution $\Phi$ such that
$\pi\circ\Phi=a\varphi$  for some $a\in G^{\C}$
and
$\Phi(\la)=\sum_{\al=0}^{m}T_{\al}\la^{\al}$ (for some $m$).
The  least such integer $m$ is called the {\it minimal uniton number} of
$\varphi$ (or of $\Phi$). The next fundamental result is that any harmonic
map which admits a corresponding real extended solution has finite uniton
number:

\proclaim{Theorem 1.2 (\cite{Uh})}
Assume that $M$ is compact.
Let $\Phi : M\longrightarrow \Omega{U_n}$ be an extended solution.
Then there exists a loop $\ga\in\Omega{U_n}$
and a non-negative integer $m\leq{n-1}$ such that
\noindent
(i)
 $\ga\Phi(\la)=
\sum_{\al=0}^{m}T_{\al}\la^{\al}$,
\noindent
(ii)
$\text{Span}\{\Im\,T_0(z){\ }\vert{\ }z\in{M}\}
={\C}^{n}$.
Here $m$ is equal to the minimal uniton number of $\Phi_{-1}$.
\qed
\endproclaim

\noindent We shall refer to property (ii) as the {\it Uhlenbeck
normalization}.

Now we discuss the group action studied by Uhlenbeck, and its
generalizations.  The idea of a \ll dressing action\rr
(see, for example, \cite{ZM},\cite{ZS},\cite{Uh},\cite{BG}) is as follows. Let
$\Cal G$ be a
group and ${\Cal G}_1, {\Cal G}_2$ two subgroups of $\Cal G$ with $\Cal
G=\Cal G_1\Cal G_2$ and ${\Cal G}_1\cap{\Cal G}_2=\{e\}$, where $e$ is
the identity element of $\Cal G$.  For any $g\in {\Cal G}$, we have a unique
decomposition
$g=g_1g_2$, $g_1\in{\Cal G}_1,g_2\in{\Cal G}_2$. For $g,h\in\Cal G$,
define $g^\sharp h$ by $g^{\sharp}h=gh(h^{-1}gh)_2^{-1}=h(h^{-1}gh)_1$.
If $g,g',h\in\Cal G$, then we have
$g^{\sharp}(g^{\prime\sharp}h)=(gg^{\prime})^{\sharp}h$, so this defines an
action of $\Cal G$ on itself.

Let $T^{\C}$ be the complexification of a maximal torus $T$ of $G$.
Let
$U_+=\{\la\in S^2{\ }\vert{\ }\vert\la\vert<1\}$
and
$U_-=\{\la\in S^2{\ }\vert{\ }\vert\la\vert>1\}$
in the Riemann sphere $S^2={\C}\cup\{\infty\}$.
Set
$$
\align
\Lambda{\gc}&=
\{\ga:S^1\longrightarrow \gc\ \vert\ \ga\ \text{smooth}\},
\\ \Lambda_{+}{\gc}&=
\{\ga\in\Lambda{\gc}{\ }\vert{\ }\ga
\text{ extends continuously to a holomorphic map }
U_+\longrightarrow \gc\}, \\
\Lambda_{-}{\gc}&=
\{\ga\in\Lambda{\gc}{\ }\vert{\ }\ga
\text{ extends continuously to a holomorphic map }
U_-\longrightarrow \gc\}, \\
\Lambda_{-}^\ast{\gc}&=
\{\ga\in\Lambda_{-}{\gc}{\ }\vert{\ }\ga(1)=e\},\\
\Delta\gc&=
\{\delta\in\Lambda{\gc}{\ }\vert{\ }
\delta : S^{1}\longrightarrow T^{\C}\subseteq \gc
\text{ is a homomorphism }\}.
\endalign
$$
The following fact is known as the Birkhoff decomposition (\cite{PS}): the
map
$$
\Lambda_{-}\gc\times\Delta\gc\times\Lambda_{+}\gc\longrightarrow
\Lambda{\gc},\quad
(\ga_{-},\delta,\ga_{+})\longrightarrow
\ga_{-}\delta\ga_{+}
$$
is surjective.
Moreover,  $\Lambda_{-}^{\ast}\gc\times\Lambda_{+}\gc$ maps
diffeomorphically to $\Lambda_{-}\gc\Lambda_{+}\gc$, which is an open
dense subset of the identity component of $\Lambda{\gc}$.
We shall now take $\Cal G=\flgc$, $\Cal G_1=\Lam^\ast\gc$, $\Cal
G_2=\Lap\gc$ in the definition of dressing action. Since $\Cal G_1\Cal G_2$
is not quite equal to $\Cal G$ here, we use the term
\lq\lq pseudo-action\rq\rq:

\noindent{\it Definition.} The {\it Birkhoff pseudo-action} of
$\Lambda{\gc}$ on itself is defined by $\ga^{\sharp}\delta=
\ga\delta(\delta^{-1}\ga\delta)_{+}^{-1}=\de(\de^{-1}\ga\de)_{-}
\in\Lambda{\gc}$, for $\ga,\de\in\flgc$ with
$\delta^{-1}\ga\delta\in\Lambda_{-}^{\ast}\gc\Lambda_{+}\gc$.

We can also consider \lq\lq{generalized Birkhoff pseudo-actions\rq\rq\
(\cite{BG}). Let $C_1,\dots,C_k$ be oriented
circles of radius $r$ on the Riemann sphere $S^2={\C}\cup\{\infty\}$.
Let $I_i$ and $E_i$ denote the interior and exterior of $C_i$ for each
$i=1,\dots,k$.
Set $C=C_1\cup\dots\cup C_k$, $I=I_1\cup\dots\cup I_k$
and $E=E_1\cap\dots\cap E_k$.
We assume in addition that $\bar I_i\cap\bar I_j=\emptyset$
for $i\ne j$ and $1\in E$.
Let
$$
\Lambda^{1,\dots,k}\gc
=\{\ga : C\longrightarrow\gc\ \vert\ \ga\text{ smooth } \},
$$
which is isomorphic to a direct product of $k$ copies of
$\Lambda{\gc}$.
Set
$$
\align
\Lambda_E{\gc}
&=\{\ga\in\Lambda^{1,\dots,k}\gc{\ }\vert{\ }
\ga \text{ extends continuously to a holomorphic map }
E\longrightarrow \gc\},
\\
\Lambda_I{\gc}
&=\{\ga\in\Lambda^{1,\dots,k}\gc{\ }\vert{\ }
\ga \text{ extends continuously to a holomorphic map }
I\longrightarrow \gc\},
\\
\Lambda_E^{\ast}{\gc}
&=\{\ga\in\Lambda_{E}\gc{\ }\vert{\ }
\ga(1) =e\},
\\
\Delta^{1,\dots,k}\gc
&=\{\delta\in\Lambda^{1,\dots,k}\gc{\ }\vert
{\ }\delta : C\longrightarrow T^{\C}\subseteq \gc\ \text{is a
homomorphism}\}.
\endalign
$$
(To say that the map $\de$ is a homomorphism means that it can be written
in the form
$\delta(\la)=
(\{(\la-c_{i})/r\}^{b_1},\dots,
\{(\la-c_{i})/r\}^{b_n})$
for
$\la \in C_{i}=\{\la\in S^2\ \vert\ \vert\la-c_i\vert=r\}$.)
There is an analogue of the Birkhoff decomposition in this situation, namely
(see \cite{BG}):
$\Lambda^{1,\dots,k}\gc=\Lambda_E\gc\Delta\gc\Lambda_I\gc$.
Moreover, under the multiplication map,
$\Lambda_E^{\ast}\gc\times\Lambda_I\gc$ is diffeomorphic to
$\Lambda_E\gc\Lambda_I\gc$,
which is an open dense subset of the identity component of
$\Lambda^{1,\dots,k}\gc$. If we take $\Cal G=\La^{1,\dots,k}\gc$, $\Cal
G_1=\La_E^\ast\gc$, $\Cal G_2=\La_I\gc$ in the definition of a dressing
action, we obtain:

\noindent{\it Definition.}
The {\it generalized Birkhoff pseudo-action} of
$\La^{1,\dots,k}\gc$ on itself is defined by $\ga^{\sharp}\delta=
\ga\delta(\delta^{-1}\ga\delta)_I^{-1}=\delta(\delta^{-1}\ga\delta)_E\in
\Lambda^{1,\dots,k}\gc$, for $\ga,\delta\in\Lambda^{1,\dots,k}\gc$
with $\delta^{-1}\ga\delta\in\Lambda_E^{\ast}\gc\Lambda_I\gc$.

The main reason for studying such pseudo-actions is:

\proclaim {Proposition 1.3 (\cite{ZM},\cite{ZS},\cite{Uh},\cite{BG})}
Let $g\in\Lambda^{1,\dots,k}\gc$
and let $\Phi$ be an extended solution.
If $\Phi^{-1}(z)g\Phi(z)\in\Lambda_E^{\ast}\gc\Lambda_I\gc$ for each
$z\in M$, then the map $g^\sharp{\Phi}$
is also an extended solution.
\qed
\endproclaim

\noindent (We assume that $\Phi_{\la}$ is defined for all $\la$ in some
region which includes $C$. For example, this is the case if $C$ does not
contain the points $0,\infty$ and if we choose $\sigma\equiv e$ in Theorem
1.1.) The pseudo-action of $\Lambda^{1,\dots,k}\gc$ on extended solutions
gives rise to a pseudo-action on harmonic maps, by means of the formula
$g^\sharp(\pi\circ\Phi)=\pi\circ g^\sharp\Phi$.
This is not quite
well-defined, as the extended solution $\Phi$ corresponding to a harmonic
map
$M\longrightarrow G$
is determined only up to left translation in $\blg$.
However, the non-uniqueness will be of no consequence in this article.

Let us impose now the following \lq\lq reality conditions\rq\rq:
(1) the equator $S^{1}$ is contained in $E$,
(2) $0,\infty\in I$, and
(3) $C=C_1\cup\dots\cup C_k$ is preserved by the transformation
$\la\longrightarrow \bar{\la}^{-1}$.
We call an element $g\in \Lambda^{1,\dots,k}\gc$
{\it real} if $g(\bar\la^{-1})^{\ast}=g(\la)^{-1}$ for each
$\la\in C$. It is easy to check that $g^\sharp\Phi$ is a real extended
solution if $g$ and $\Phi$ are real.
We denote by $\Lambda_{\R}^{1,\dots,k}\gc$ the
subgroup of real elements of $\Lambda^{1,\dots,k}\gc$, and by
$\Lambda_{E,\R}\gc,\Lambda_{E,\R}^{\ast}\gc,\Lambda_{I,\R}\gc,
\Delta_{\R}\gc$
the subgroups of real elements of
$\Lambda_E\gc,\Lambda_E^{\ast}\gc$,{}$\Lambda_I\gc,\Delta\gc$.

We shall now give a contour integral expression for the generalized
Birkhoff pseudo-action of
${\Lambda}^{1,\dots,k}\gc$ on ${\Lambda}_{E}\gc$. Note that for
$\de\in\La_E\gc$ the formula for $\ga^\sharp\de$ simplifies to
$\ga^\sharp\de=\ga\de(\ga\de)_I^{-1}=(\ga\de)_E$.

\proclaim {Lemma 1.4}
 Let
$g \in {\Lambda}^{1,\dots,k}\gc$ and $h \in {\Lambda}_{E}\gc$.
Assume that $h^{-1}gh \in {\Lambda}_{E}^{\ast}\gc{\Lambda}_{I}\gc$, so
that $g^{\sharp}h \in {\Lambda}_{E}^{\ast}\gc$ is well-defined.
Then
$$
(g^{\sharp}h)(\la)-h(\la)
={\la-1 \over 2\pi i}
\int_C
{h(\la)h^{-1}(\mu)(g^{-1}(\mu)-e)(g^{\sharp}h)(\mu)
\over
(\mu-1)(\mu-\la)}
d\mu
$$
for each $\la \in E$.
\endproclaim

\demo {Proof}  By using Cauchy's Integral Theorem, we obtain
$$
(h^{-1}gh)_E(\la)-e={\la-1\over2\pi i}\int_C{((h^{-1}gh)^{-1}(\mu)-
e)(h^{-1}gh)_E(\mu)\over(\mu-1)(\mu-\la)}d\mu.
$$
Multiplying by $h(\la)$ on the left, we obtain the required formula.
\qed\enddemo

Using this lemma, we derive a formula for the
infinitesimal action of ${\Lambda}^{1,\dots,k}\gc$ on ${\Lambda}_{E}\gc$.
Let $\{g_t\}_{\vert t\vert<\ep}$ be a curve in ${\Lambda}^{1,\dots,k}\gc$
with $g_0=e$ and set
$V={d\over dt}g_t\vert_{t=0} \in {\Lambda}^{1,\dots,k}{\frak g}^{\C}$.
Let $h \in \Lambda_E\gc$. Note that for each $t$ sufficiently close to $0$,
$h^{-1}g_{t}h \in {\Lambda}_{E}^{\ast}\gc{\Lambda}_{I}\gc$ and hence
$g_t^{\sharp}h \in {\Lambda}_{E}^{\ast}\gc$ is defined. Set
$$
V^{\sharp}_h={d \over dt}g_t^{\sharp}h\big|_{t=0}
\in T_h\Lambda_E\gc.
$$

\proclaim{Proposition 1.5}
For each $\la \in E$, we have
$$
dL_h^{-1}(V^{\sharp}_h)(\la)
=-{\la-1 \over 2\pi i}
\int_C
{h^{-1}(\mu)V(\mu)h(\mu) \over (\mu-1)(\mu-\la)}
d\mu.
$$
Here $L_h$ denotes left translation by $h$ in the
group $\Lambda_E\gc$.
\endproclaim

\demo {Proof} Replace $g$ by $g_t$ in the formula of Lemma 1.4.
By differentiating at $t=0$, we obtain the required formula.
\qed
\enddemo

 \proclaim{Corollary 1.6}
Assume that $0 \in I_1, {\infty} \in I_2$.
If $g\in \Lambda_I\gc$ satisfies $g\vert_{I_i}=e$ for $i=1,2$ and $h\in
\Lambda_E\gc$ extends to a holomorphic map
${\C}^{\ast}=S^2\setminus\{0,\infty\}\longrightarrow \gc$,
then $g^{\sharp}h$ exists and $g^{\sharp}h=h$.\qed
\endproclaim

\noindent Thus, if $\Phi$ is a real extended solution, which without loss of
generality we may assume is defined for all $\la\in\CS$, then it is only
necessary to consider generalized Birkhoff pseudo-actions with $C=C_1\cup
C_2$, where $C_1,C_2$ are circles around $0,\infty$ respectively.

$${}$$

\heading
\S{2}. Properties of the Uhlenbeck pseudo-action
\endheading

In this section we shall study the pseudo-action introduced
by Uhlenbeck in \cite{Uh}. It can be regarded as the generalized Birkhoff
pseudo-action given by the choice  of circles
$$
C_0^{\ep}=\{\la\in S^2\ \vert\
\vert\la\vert=\ep\},\quad
C_{\infty}^{\ep}=\{\la\in S^2\ \vert\
\vert\la\vert=\frac 1{\ep}\}, $$ where $0<\ep<1$.
We shall call it the {\it Uhlenbeck pseudo-action}.
This is the simplest choice which is compatible with
the reality conditions, and by Corollary 1.6 it contains the essential
features of all the other choices.

We shall write
$\Lambda^{\ep}\gc$ for $\Lambda^{1,2}\gc$,
where $C_1=C_0^{\ep}, C_2=C_\infty^{\ep}$.
Using the notation of the previous section, we have
$C=C_{1}\bigcup C_{2}$, $I=I_1\bigcup I_{2}$ and
$E=S^{2}\setminus C\bigcup I_1\bigcup I_{2}$,
where
$$
I_{1}=\{\la \in S^2 \ \vert\  \vert\la\vert<\ep\},\quad
I_{2}=\{\la \in S^2 \ \vert\
\vert\la\vert>\frac 1{\ep}\}.
$$
We have subgroups
$\Lambda_E\gc,\Lambda_E^{\ast}\gc,\Lambda_I\gc$
of $\Lambda^{\ep}\gc$ as in the previous section. We denote by
$\Lambda_{\R}^{\ep}\gc$ the subgroup of all real elements $\ga$ of
$\Lambda^{\ep}\gc$, namely elements
 satisfying the reality condition
$\ga(\bar{\la}^{-1})^{\ast}=\ga(\la)^{-1}$ on $C$.

Let
$$\align
&{\Cal  G}=\{g:U\longrightarrow\gc\ \vert\ g\ \text{holomorphic in some
neighbourhood}\ U\ \text{of}\ \{0,\infty\}\},\\
&{\Cal  G}_{\R}=\{g\in\Cal G\ \vert\ g(\bar\la^{-1})^{\ast}=
g(\la)^{-1}\ \text{for all}\ \la\}.
\endalign
$$
Note that ${\Cal  G}$ and ${\Cal  G}_{\R}$ are connected. Let
$$
\align
&{\Cal A}=\{g\in{\Cal G}{\ }\vert{\ }\text{$g$ extends to a
$\gc$-valued rational function on $S^2$ }\},
\\
&{\Cal A}_{\R}=\{g\in{\Cal A}{\ }\vert{\ }
g(\bar\la^{-1})^{\ast}=g(\la)^{-1}
\ \text{ for all }\ \la\}.
\endalign
$$
For each $\ep$ with $0<\ep<1$, we consider
$\Lambda_I\gc$ and $\Lambda_{I,\R}\gc$ as subgroups of
${\Cal  G}$ and ${\Cal  G}_{\R}$, respectively.
We then have
$$
\bigcup_{0<\ep<1}\Lambda_{I}\gc={\Cal  G},\quad
\bigcup_{0<\ep<1}\Lambda_{I,\R}\gc={\Cal  G}_{\R}.
$$

Denote by $Lie({\Cal  G})$ and $Lie({\Cal  G}_{\R})$ the Lie algebras of
${\Cal  G}$ and ${\Cal  G}_{\R}$, respectively. For each integer $k\geq 0$
or $k=\infty$, let
$$\eqalign{
Lie(\Cal  G)_k=\{{\ }V\in Lie({\Cal  G}) {\ }\vert{\ }
&V(\la)=
\sum_{\al\geq k}V_{\al}^{(0)}\la^{\al}
\hbox{ around } 0,\cr
&V(\la)=
\sum_{\al\geq k}V_{-\al}^{(\infty)}\la^{-\al}
\hbox{ around } \infty \}.\cr}
$$
Then $Lie(\Cal G)_k$ is an ideal of $Lie({\Cal  G})$ and
$Lie(\Cal G)_k\subseteq Lie(\Cal G)_{k-1}$,
$Lie(\Cal G)_0=Lie(\Cal G)$.
Let ${\Cal  G}_k$ be the analytic subgroup of ${\Cal  G}$ generated by
the Lie algebra $Lie(\Cal G)_k$, which is a connected closed normal
subgroup of ${\Cal  G}$. (Thus, $Lie(\Cal G_k)=Lie(\Cal G)_k$.)
The quotient complex Lie algebra $Lie({\Cal
G})/Lie(\Cal G)_k$ has complex dimension $2k\dim_{\C}{\frak g}^{\C}$. We
have a sequence of surjective Lie group homomorphisms :
${\Cal  G}/{\Cal  G}_k\longrightarrow
{\Cal  G}/{\Cal  G}_{k-1}{\ }(k=1,2,\dots)$.
Set
$Lie(\Cal G)_{k,{\R}}= Lie(\Cal G)_{\R}\cap Lie(\Cal G)_{k}$,
which is a real Lie algebra. The Lie algebra $Lie(\Cal G)_{k,{\R}}$ generates
an analytic subgroup ${\Cal  G}_{k,{\R}}$ of ${\Cal  G}_{\R}$, which is a
connected closed normal subgroup of ${\Cal  G}_{\R}$. The quotient real Lie
algebra $Lie({\Cal  G}_{\R})/Lie(\Cal G)_{k,{\R}}$ has real dimension
$2k\dim{\frak g}$.

For each integer $k\geq 0$ or $k=\infty$, we set
${\Cal A}_k={\Cal A}\cap{\Cal  G}_{k}$,
${\Cal A}_{k,{\R}}={\Cal A}_{\R}\cap{\Cal  G}_{k}$.
Note that ${\Cal A}_k$ is a closed normal subgroup of
${\Cal A}$.

\proclaim{Proposition 2.1}
(i) For each $k$ with $0\leq{k}<\infty$,
the natural injective homomorphism of
${\Cal A}$
into ${\Cal  G}$ induces a Lie group isomorphism of
${\Cal A}/{\Cal A}_{k}$
onto ${\Cal  G}/{\Cal  G}_{k}$.
(ii) For each $k$ with $0\leq{k}<\infty$, the natural injective
homomorphism of ${\Cal A}_{\R}$
into ${\Cal  G}_{\R}$ induces a Lie group isomorphism of
${\Cal A}_{\R}/{\Cal A}_{k,{\R}}$
onto ${\Cal  G}_{\R}/{\Cal  G}_{k,{\R}}$.
\endproclaim

{\it Proof.}  Denote by $\sigma$ and $d\sigma$ the Lie group
homomorphism ${\Cal A}/{\Cal A}_{k}
\longrightarrow{\Cal  G}/{\Cal  G}_{k}$ and its derivative, respectively.
We have only to show that $\sigma$ is surjective.
Let $V$ be any element of $Lie({\Cal  G})$.
We take the Taylor expansions of $V$ around $0$ and $\infty$:
$V(\la)=
\sum_{\al\geq 0}V_{\al}^{(0)}\la^{\al}
\hbox{ around } 0$, and
$V(\la)=
\sum_{\al\geq 0}V_{-\al}^{(\infty)}\la^{-\al}
\hbox{ around } \infty$. By the method of indeterminate coefficients, we can
find $U\in Lie({\Cal A})$ such that $U(\la)=
\sum_{\al=0}^{k-1}V_{\al}^{(0)}\la^{\al}+
\sum_{\al=k}^{\infty}U_{\al}^{(0)}\la^{\al}$ around $0$
and $U(\la)=
\sum_{\al=0}^{k-1}V_{-\al}^{(\infty)}\la^{-\al}+
\sum_{\al=k}^{\infty}U_{-\al}^{(\infty)}\la^{-\al}$
around
$\infty$.
Hence $U-V\in Lie({\Cal A}_k)$, namely $U\equiv{V}$ mod
$Lie({\Cal A}_k)$. Thus $d\sigma$ is surjective. Since
${\Cal  G}/{\Cal  G}_{k}$ is connected, $\sigma$ is also surjective.
This proves (i). The proof of (ii) is similar.
\qed

For each integer $k\geq 0$ or $k=\infty$, let
$$\align
&{\Cal X}_k=\{\ga : {\C}^{\ast}\longrightarrow \gc\ \vert\ \ga\
\text{holomorphic,}\ \ga(1)=e,\\
&\quad\quad\quad\quad\text{and}\ \ga(\la)=
\sum_{\vert\al\vert\leq k}A_{\al}{\la}^{\al},
\ga^{-1}(\la)=
\sum_{\vert\al\vert\leq k}B_{\al}{\la}^{\al}\}\\
&{\Cal X}_{k,{\R}}=\{\ga\in\Cal X_k\ \vert\ \ga(\bar\la^{-
1})^\ast=\ga(\la)^{-1}\ \text{for all}\ \la\}.
\endalign
$$
Similary, let
$$\align
&{\Cal X}_k^{+}=\{\ga : {\C}^{\ast}\longrightarrow \gc\ \vert\ \ga\
\text{holomorphic,}\ \ga(1)=e,\\
&\quad\quad\quad\quad\text{and}\ \ga(\la)=
\sum_{\al=0}^kA_{\al}{\la}^{\al},
\ga^{-1}(\la)=
\sum_{\al=0}^kB_{-\al}{\la}^{-\al}\}\\
&{\Cal X}_{k,{\R}}^{+}=\{\ga\in\Cal X_k^{+}\ \vert\
\ga(\bar\la^{-1})^{\ast}=\ga(\la)^{-1}\ \text{for all}\ \la\}.
\endalign
$$
We can consider
${\Cal X}_{k,{\R}}$
and
${\Cal X}_{k,{\R}}^{+}$
as subspaces of $\Omega G$.
Set
${\Cal X}={\Cal X}_{\infty}$
and
${\Cal X}_{\R}={\Cal X}_{\infty,{\R}}$.
The point of these definitions is that a harmonic map of finite uniton
number gives rise to an extended solution with values in
${\Cal X}_{k,{\R}}^{+}$, for some $k$.

Uhlenbeck obtained the following theorem by showing that any
element of ${\Cal A}_{\R}$ decomposes into a product of elements of
\lq\lq{simplest type}\rq\rq, then by showing that the action is defined
for any element of simplest type.  See also \cite{Be}.

\proclaim{Theorem 2.2 (\cite{Uh})}
For each $g\in {\Cal A}_{\R}$ and each $\ga\in {\Cal X}_{\R}$,
$g^{\sharp}\ga\in {\Cal X}_{\R}$ is well-defined.\qed
\endproclaim

We call the action of ${\Cal A}_{\R}$ on ${\Cal X}_{\R}$
{\it the Uhlenbeck action}.

\proclaim{Theorem 2.3}
(i) If $V\in Lie(\Cal G)_{2k}$ and $\ga\in {\Cal X}_k$,
then $V^{\sharp}_{\ga}=0$.
(ii) If $g\in {\Cal  G}_{2k}$ and ${\ga}\in{\Cal X}_k$,
then $g^{\sharp}\ga\in{\Cal X}_k$ is defined and
$g^{\sharp}\ga=\ga$.
\endproclaim

\proclaim{Theorem 2.4}(i) If $V\in Lie(\Cal G)_{k}$ and
$\ga\in {\Cal X}_k^{+}$, then $V^{\sharp}_{\ga}=0$.
(ii) If $g\in {\Cal  G}_{k}$ and ${\ga}\in{\Cal X}_k^{+}$,
then $g^{\sharp}\ga\in{\Cal X}_k$ is defined and
$g^{\sharp}\ga=\ga$.
\endproclaim

\demo {Proof of Therem 2.3}
(i)
Let
$V\in Lie(\Cal G)_{2k}$ and $\ga\in {\Cal X}_k$.
Then we have
$\ga(\la)=
\sum_{\vert\al\vert\leq k}A_{\al}\la^{\al}$
and
$\ga^{-1}(\la)=
\sum_{\vert\al\vert\leq k}B_{\al}\la^{\al}$
for $\la\in{\C}^{\ast}$.
By Proposition 1.5 we have, for $\la\in S^1$,
$$
dL_\ga^{-1}(V^{\sharp}_\ga)(\la)
=-{\la-1 \over 2\pi i}\{\int_{C_{0}}
{\ga^{-1}(\mu)V(\mu)\ga(\mu) \over (\mu-1)(\mu-\la)} d\mu+
\int_{C_{\infty}} {\ga^{-1}(\mu)V(\mu)\ga(\mu) \over (\mu-1)(\mu-\la)}
d\mu\}.
$$
Denote by (A) and (B) the first term and the second term on the
right-hand side of this formula. By  assumption we have
$$
\align
&V(\la)=
\sum_{\al\geq 2k}V_{\al}^{(0)}\la^{\al}
\hbox{ on }\bar I_1,\\
&V(\la)=
\sum_{\al\geq 2k}V_{-\al}^{(\infty)}\la^{-\al}
\hbox{ on }\bar I_2.
\endalign
$$
On the circle $C_{0}$, we have
$$
\ga^{-1}(\mu)V(\mu)\ga(\mu)
=\sum_{\vert\al\vert\leq{k},\vert\al^{\prime}\vert\leq{k},\be\geq{2k}}
B_{\al^{\prime}}V_{\be}^{(0)}
A_{\al}\mu^{\al^{\prime}+\be+\al}.
$$
Write
$$
{1 \over (\mu-1)(\mu-\la)}
=\sum_{{\al}^{\prime\prime}\geq{0}}
a_{{\al}^{\prime\prime}}^{(0)}\mu^{{\al}^{\prime\prime}}
$$
around $0$.
Then the first integrand is
$$
\sum_{{\al}^{\prime\prime}\geq{0},
\vert\al\vert\leq{k},\vert\al^{\prime}\vert\leq{k},\be\geq{2k}}
a_{{\al}^{\prime\prime}}^{(0)}
B_{\al^{\prime}}V_{\be}^{(0)}
A_{\al}
\mu^{{\al}^{\prime\prime}+\al^{\prime}+\be+\al}.
$$
Since
${\al}^{\prime\prime}+\al^{\prime}+\be+\al\geq{0}$,
in particular
${\al}^{\prime\prime}+\al^{\prime}+\be+\al
\not={-1}$, we obtain (A)$=0$.
\noindent
On the circle $C_{\infty}$, we have
$$
\ga^{-1}(\mu)V(\mu)\ga(\mu)
=\sum_{\vert\al\vert\leq{k},\vert\al^{\prime}\vert\leq{k},\be\geq{2k}}
B_{\al^{\prime}}V_{-\be}^{(\infty)}
A_{\al}\mu^{\al^{\prime}-\be+\al}.
$$
Write
$$
{1 \over (\mu-1)(\mu-\la)}
=\sum_{{\al}^{\prime\prime}\geq{2}}
a_{-{\al}^{\prime\prime}}^{(\infty)}\mu^{-{\al}^{\prime\prime}}
$$
around $\infty$.
Then the second integrand is
$$
\sum_{{\al}^{\prime\prime}\geq{2},
\vert\al\vert\leq{k},\vert\al^{\prime}\vert\leq{k},\be\geq{2k}}
a_{-{\al}^{\prime\prime}}^{(\infty)}
B_{\al^{\prime}}V_{-\be}^{(\infty)}
A_{\al}
\mu^{-{\al}^{\prime\prime}+\al^{\prime}-\be+\al}.
$$
Since
$-{\al}^{\prime\prime}+\al^{\prime}-\be+\al\leq
{-2+k-2k+k}=-2$,
in particular
$-{\al}^{\prime\prime}+\al^{\prime}-\be+\al\neq{-1}$,
we obtain (B)$=0$.

(ii)  By (i), there is a neighbourhood ${\Cal U}$ of $e$ in
${\Cal  G}_{2k}$
such that
$g^{\sharp}\ga$
exists and
$g^{\sharp}\ga=\ga$ for each ${\ga}\in{\Cal U}$.
Since the group ${\Cal  G}_{2k}$ is connected, ${\Cal  G}_{2k}$
is generated by elements of ${\Cal U}$. Hence we obtain (ii).
\qed\enddemo

\demo{Proof of Theorem 2.4}
Let
$V\in Lie(\Cal G)_{k}$ and $\ga\in {\Cal X}_k^{+}$.
Then we have
$\ga(\la)=
\sum_{\al=0}^{k}A_{\al}\la^{\al}$
and
$\ga^{-1}(\la)=
\sum_{\al=0}^{k}B_{-\al}\la^{-\al}$
for $\la\in{\C}^{\ast}$.
By assumption, we have
$$
\align
&V(\la)=
\sum_{\al\geq k}V_{\al}^{(0)}\la^{\al}
\hbox{ on }\bar I_1,\\
&V(\la)=
\sum_{\al\geq k}V_{-\al}^{(\infty)}\la^{-\al}
\hbox{ on }\bar I_2.
\endalign
$$
As in the proof of Theorem 2.3, the first integrand in the expression for
$dL_\ga^{-1}(V^{\sharp}_\ga)$ is
$$
\sum_{{\al}^{\prime\prime}\geq{0},
0\leq\al\leq{k},0\leq\al^{\prime}\leq{k},\be\geq{k}}
a_{{\al}^{\prime\prime}}^{(0)}
B_{-\al^{\prime}}V_{\be}^{(0)}
A_{\al}
\mu^{{\al}^{\prime\prime}-\al^{\prime}+\be+\al}.
$$
Since
${\al}^{\prime\prime}-\al^{\prime}+\be+\al
\geq 0-k+k+0=0$,
in particular
${\al}^{\prime\prime}-\al^{\prime}+\be+\al
\neq{-1}$,
we obtain (A)$=0$.
The second integrand is
$$
\sum_{{\al}^{\prime\prime}\geq{2},
0\leq\al\leq{k},0\leq\al^{\prime}\leq{k},\be\geq{k}}
a_{-{\al}^{\prime\prime}}^{(\infty)}
B_{-\al^{\prime}}
V_{-\be}^{(\infty)}
A_{\al}
\mu^{-{\al}^{\prime\prime}-\al^{\prime}-\be+\al}.
$$
Since
$-{\al}^{\prime\prime}-\al^{\prime}-\be+\al
\leq -2+0-k+k=-2$,
in particular
$-{\al}^{\prime\prime}-\al^{\prime}-\be+\al
\neq{-1}$,
we obtain (B)$=0$. This proves (i).
By the same argument as in the proof of Theorem 1.3,
(ii) follows from (i).
\qed\enddemo

Theorem 2.4 implies that, for each $k$ with $0\leq k<\infty$, the
pseudo-actions of the infinite dimensional Lie groups
${\Cal A}_{\R}$
and
${\Cal  G}_{\R}$
on
${\Cal X}_{k,\bold{R}}^{+}$
collapse to the pseudo-actions of the finite dimensional Lie groups
${\Cal A}_{\R}/{\Cal A}_{k,{\R}}$
and
${\Cal  G}_{\R}/{\Cal  G}_{k,{\R}}$,
respectively. Moreover, by Theorem 2.2 and Proposition 2.1, we
see that these pseudo-actions are in fact actions.
In \S{5} we shall prove by a different argument
that the pseudo-action of
${\Cal G}_{\R}/{\Cal G}_{k,{\R}}$
on
${\Cal X}_{k,\bold{R}}^{+}$
is an action, i.e. without using Theorem 2.2.

$${}$$

\heading \S 3. The natural action
\endheading

In this section we study a different group action on the space of
extended solutions $M\longrightarrow\blg$. This approach depends on
recognising
explicitly the role of the loop group $\blg$. It is well known that
$\blg$ enjoys many of the properties of a {\it finite dimensional}
generalized flag
manifold (or K\"ahler C-space); one reason for this is that $\blg$
arises as an orbit of the \ll adjoint action\rr for the Lie group
$S^1\ltimes\flg$. The semi-direct product here is defined with
respect to the action of $S^1$ on the free loop group
$\flg=\Map(S^1,G)$ by rotation of the loop parameter. (That is,
$(e^{2\pi i\varphi},\ga(e^{2\pi it}))
\cdot(e^{2\pi i\psi},\de(e^{2\pi it}))=
(e^{2 \pi i(\varphi+\psi)},\ga(e^{2\pi it})\de(e^{2\pi i(t-\varphi)}))$.)
Indeed, the isotropy subgroup of
the point $(i,0)\in i\R\ltimes\flgg$ is the group $S^1\times G$,  so
$$
\blg\cong\frac{\flg}{G}\cong\frac{S^1\ltimes\flg}{S^1\times G}.
$$
The analogy can be strengthened by introducing the \ll Grassmannian
model\rr of $\blg$ (see \cite{PS}, Chapters 7,8). This is a submanifold of
an infinite dimensional Grassmannian on which $S^1\ltimes\flg$
acts transitively, with isotropy subgroup $S^1\times G$, and it
provides a geometrical basis for the above identification. We shall
review briefly this construction.

Let $e_1,\dots,e_n$ be an orthonormal basis of $\Cn$. Let $H^{(n)}$ be
the Hilbert space $L^2(S^1,\Cn)= \allowmathbreak
\Span\{\la^ie_j\ \vert\ i\in\Z,\ j=1,\dots,n\}$, and let $H_+$ be the
subspace $\Span\{\la^ie_j\ \vert\ i\ge0,\ j=1,\dots,n\}$. The group
$\blu$ acts naturally on $H^{(n)}$ by multiplication, and we have a map
from $\blu$ to the Grassmannian of all closed linear subspaces of
$H^{(n)}$, given by $\ga\longmapsto\ga H_+=\{\ga f\ \vert\ f\in H_+\}$. It
is easy to see that this map is injective. Regarding the image, one
has:

\proclaim{Theorem 3.1 (\cite{PS})}
The image $Gr_{\infty}^{(n)}$ of the map
$\ga\longmapsto\ga H_+$ consists of all closed linear subspaces $W$ of
$H^{(n)}$ which satisfy
\roster
\item $\la W\subseteq W$,
\item the orthogonal projections $W\longrightarrow H_+$ and
$W\longrightarrow (H_+)^\perp$
are respectively Fredholm and Hilbert-Schmidt, and
\item the images of the orthogonal projections $W^\perp\longrightarrow
H_+$
and $W\longrightarrow (H_+)^\perp$ consist of smooth functions.
\endroster

\noindent Moreover, if $\ga\in\blu$ and $W=\ga H_+$, then
$\deg(\det\ga)$ is minus the index of the orthogonal projection
operator $W\longrightarrow H_+$.\qed
\endproclaim

\noindent This is the Grassmannian model of $\blu$.

Now suppose $G$ is a compact connected Lie group with trivial
centre. Via the adjoint representation, we may consider $G$ as a
subgroup of $U_n$ (where $n=\dim\,G$) and $\blg$ as a subgroup of
$\blu$. The Hilbert space $H^{(n)}$ inherits the structure of a Lie
algebra from ${\frak g}^{\bold C}$, and its Hermitian inner product arises
from
the Killing form of $\frak g$.

\proclaim{Corollary 3.2 (\cite{PS})}
The image of $\blg$ under the map
$\ga\longmapsto\ga H_+$ consists of all closed linear subspaces $W$ of
$H^{(n)}$ which satisfy
\roster
\item $\la W\subseteq W$,
\item the orthogonal projections $W\longrightarrow H_+$ and
$W\longrightarrow (H_+)^\perp$
are respectively Fredholm and Hilbert-Schmidt, and
\item $W^{sm}$ is a subalgebra of the Lie algebra $H^{(n)}$, where
$W^{sm}$ is the space of smooth functions in $W$, and
\item $\overline W^\perp=\la W$.\qed
\endroster
\endproclaim

\noindent This is the Grassmannian model of $\blg$. If $G^\prime$ is
any locally isomorphic group, we can obtain a Grassmannian model
for $\Omega G^\prime$, because it suffices to give a model for the
identity component, and the identity components of $\blg$ and
$\Omega G^\prime$ may be identified. In particular, this shows that
one has a Grassmannian model for any compact semisimple Lie group.

The complexified group $\flgc$ also acts transitively on the
Grassmannian model, with isotropy subgroup $\flpgc$ at $H_+$.
Hence one obtains the identification
$$
\blg\cong\frac{\flgc}{\flpgc}.
$$
It follows that $\flgc=\blg\cdot\flpgc$. Since
$\blg\cap\flpgc=\{e\}$, we have a factorization theorem: any
$\ga\in\flgc$ can be written as $\ga=\ga_u\ga_+$, where
$\ga_u,\ga_+$ are uniquely defined elements of $\blg,\flpgc$
respectively.
If $\ga\in\flgc$, we shall write $[\ga]$ for the coset
$\ga(\flpgc)\in\flgc/\flpgc\cong\blg$.
Thus, the natural action of
$\flgc$ on $\blg$, denoted by the symbol $\natural$, may be written
$$
\ga\nat\de=[\ga\de]=(\ga\de)_u.
$$

\noindent{\it Definition.}
Let $\Phi:M\longrightarrow\blg$ be an extended
solution. Let $\ga\in\flgc$. We define the {\it
natural action} of $\ga$ on $\Phi$ by
$\ga\nat\Phi=[\ga\Phi]=(\ga\Phi)_u$.

Let $\Phi:M\longrightarrow\blg$ be a smooth map. By the Grassmannian
model,
this may be identified with a map $W:M\longrightarrow Gr_\infty^G$, where
$W(z)=\Phi(z)H_+$. The extended solution equations for $\Phi$ are
equivalent to the conditions
$$\gather
\ddzb \sec W\subseteq \sec W\tag 1\\
\ddz \sec W\subseteq\sec\la^{-1}W\tag 2
\endgather
$$
where $\sec W$ denotes the space of (locally defined) smooth maps
$f:M\longrightarrow H^{(n)}$ with $f(z)\in W(z)$ for all $z$. The first
condition is
simply the condition that $\Phi$ be holomorphic. The second
condition is a horizontality condition on the derivative of $\Phi$
(this terminology will be explained in the next section).

\proclaim{Proposition 3.3}
Let $\Phi:M\longrightarrow\blg$ be an extended
solution. Let $\ga\in\flgc$. Then $\ga\nat\Phi$
is also an extended solution.
\endproclaim

\demo{Proof} Let $W:M\longrightarrow Gr_\infty^G$ be the map
corresponding to
$\Phi$; thus $\ga W$ corresponds to $\ga\nat\Phi$. If $W$ satisfies
equations (1) and (2), then so does $\ga W$, as multiplication by
$\ga$ commutes with the differentiation with respect to $z$ or $\bar z$
and with multiplication by $\la^{-1}$.
\qed\enddemo

To understand this action, it is helpful to consider the following
concrete examples. We shall show later that these examples represent
special cases of the action.

\noindent{\bf Example 3.4} Let $\varphi:M\longrightarrow U_n$ be a
harmonic map
with minimal uniton number $1$. Then  $\varphi=\pi\circ \Phi$, where
$\Phi:M\longrightarrow\grkcn$ is a holomorphic map (for some $k$), and
where
$\pi:\grkcn\longrightarrow U_n$ is a totally geodesic embedding.
More explicitly, there exists some $a\in U_n$ such that
$\varphi(z)=a(\pi_{\Phi(z)}-\pi^\perp_{\Phi(z)})$, where $\pi_{\Phi(z)}$
denotes the
orthogonal projection $\C^n\longrightarrow \Phi(z)$ with respect to the
Hermitian
inner product of $\C^n$. The embedding $\pi:\grkcn\longrightarrow U_n$ is
then
given by $V\longmapsto a(\pi_V-\pi^\perp_V)$. Conversely, any map
$\varphi$ of this form (with $\Phi$ non-constant) is a harmonic map with
minimal uniton number $1$. Since the standard action of the complex
group $\glnc=\text{Aut}(\C^n)$ on $\grkcn$ is holomorphic, we
obtain an action of $\glnc$ on holomorphic maps
$M\longrightarrow\grkcn$.
Thus, an element $A$ of $\glnc$ gives rise to a new holomorphic map
$A\nat\varphi=\pi(A\Phi)$.

\noindent{\bf Example 3.5}
It is well known (see \cite{EL}) that all harmonic maps
$\varphi:S^2\longrightarrow\CP^{n}$
are of the form $\varphi=\pi\circ\Phi$, where $\Phi:S^2\longrightarrow
F_{r,r+1}(\C^{n+1})$ is (a) holomorphic with respect to the natural
complex structure of $F_{r,r+1}(\C^{n+1})$, and (b) horizontal with
respect to the projection $\pi:F_{r,r+1}(\C^{n+1})\longrightarrow\CP^{n}$.
Here,
$F_{r,r+1}(\C^{n+1})$ is the space of flags of the form $\{0\}\subseteq
E_r\subseteq E_{r+1}\subseteq\C^{n+1}$. Conversely, given a holomorphic
horizontal map $\Phi$, the map $\varphi=\pi\circ\Phi$ is harmonic.
If the flag corresponding to $\Phi(z)$ is denoted by $\{0\}\subseteq
W_r(z)\subseteq W_{r+1}(z)\subseteq\C^{n+1}$, then the holomorphicity and
horizontality conditions are
$$\gather
\ddzb \sec W_i\subseteq \sec W_i,\ i=r,r+1\tag 1\\
\ddz \sec W_r\subseteq \sec W_{r+1}.\tag 2
\endgather
$$
The standard action of $GL_{n+1}(\C)$ on $F_{r,r+1}(\C^{n+1})$
preserves both these conditions because of the linearity of the
derivative. Hence for any $A\in GL_{n+1}(\C)$, we obtain a new
harmonic map $A\nat\varphi=\pi(A\Phi)$. This action of $GL_{n+1}(\C)$
on harmonic maps $S^2\longrightarrow\CP^{n}$ was studied in \cite{Gu}.

More generally, if $M$ is a Riemann surface, complex isotropic harmonic
maps $\varphi:M\longrightarrow \CP^n$ correspond to holomorphic
horizontal maps $\Phi: M\longrightarrow F_{r,r+1}(\C^{n+1})$. Thus, we
obtain an action of $GL_{n+1}(\C)$ on complex isotropic harmonic maps.

\noindent{\bf Example 3.6} There is a similar description of
harmonic maps from $S^2$ to $S^n$ or $\R P^n$. It suffices to
consider harmonic maps $\varphi:S^2\longrightarrow\R P^{2n}$,
as the other cases
can be deduced from this one. Let $Z_n$ be the space of (complex)
$n$-dimensional subspaces $V$ of $\C^{2n+1}$ such that $V$ and
$\overline V$ are orthogonal with respect to the standard Hermitian
inner product of $\C^{2n+1}$, i.e. such that $V$ is \ll isotropic\rrr.
There is a projection map $\pi:Z_n\longrightarrow\R P^{2n}$, which
associates to
$V$ the $(+1)$-eigenspace of the operator $x\longmapsto\bar x$ on
$(V\oplus\overline V)^\perp$.  It is known (see
\cite{Ca1},\cite{Ca2},\cite{Ba}) that such
harmonic maps are of the form $\varphi=\pi\circ\Phi$ where
$\Phi:S^2\longrightarrow Z_n$ is a holomorphic map which is horizontal
with respect to $\pi$.
The holomorphicity and horizontality conditions are
$$\gather
\ddzb\sec\Phi\subseteq\sec\Phi\tag 1\\
\ddz\sec\Phi\perp\sec\overline{\Phi}.\tag 2
\endgather
$$
The standard action of $SO^{\C}_{2n+1}$ on $Z_n$ preserves both
these conditions, hence we obtain an action of $SO^{\C}_{2n+1}$ on
harmonic maps.

More generally, if $M$ is a Riemann surface, isotropic harmonic maps from
$M$ into $S^n$ or $\R P^n$  correspond to holomorphic horizontal maps $\Phi:
M\longrightarrow Z_n$, and we obtain an action of $SO^{\C}_{2n+1}$ on such
maps.

The harmonic maps arising in these three examples fit into a more general
framework, described in \cite{Br}, \cite{BR}, which we shall recall briefly.
Let $G/H$ be a generalized flag manifold, i.e. the orbit of a point $P$ of
$\lieg$ under the adjoint representation. It is well known that the complex
group $\gc$ acts transitively on $G/H$. If $G_P$ is the isotropy subgroup at
$P$, then we have an identification $G/H\cong \gc/G_P$. This endows $G/H$
with a complex structure, and the holomorphic tangent bundle of $G/H$ may
be identified with the homogeneous bundle
$\gc\times_{G_P}(\liegc/\liegp)$.

Without essential loss of generality (see \cite{BR}) we may assume that the
linear endomorphism $\ad P$ on $\liegc$ has eigenvalues in $i\Z$. If the
$(i\ell)$-eigenspace is denoted by $\lieg_\ell$, then one has
$\lieg_0=\liehc$,
$\liegp=\bigoplus_{i\le 0}\lieg_i$, and
$[\lieg_i,\lieg_j]\subseteq\lieg_{i+j}$. Let $\liekc=\bigoplus_{i\
\text{even}}\lieg_i$. Then $(\liegc,\liekc)$ is a symmetric pair, and (up to
local isomorphism) one obtains a symmetric space $G/K$, where $K=\{g\in
G\ \vert\ g(\exp\pi P)=(\exp\pi P)g\}$.

The natural map $\pi:G/H\longrightarrow G/K$ is a \lq\lq twistor
fibration\rq\rq; it gives rise to a relation between harmonic maps
$M\longrightarrow G/K$ and holomorphic maps $M\longrightarrow G/H$. The
simplest aspect of this relation may be expressed in terms of the {\it
super-horizontal distribution}, which is by definition the holomorphic
subbundle $\gc\times_{G_P}(\liegp\oplus\lieg_1/\liegp)$ of
$\gc\times_{G_P}(\liegc/\liegp)\cong T_{1,0}G/H$. A holomorphic map
$\Phi:M\longrightarrow G/H$ is said to be super-horizontal if it is
tangential to the super-horizontal distribution. It is shown in
\cite{Br},\cite{BR} that:

\noindent ($\dagger$) If $\Phi$ is holomorphic and super-horizontal, then
$\varphi=\pi\circ\Phi$ is harmonic.

\noindent Clearly the action of $\gc$ preserves holomorphicity and
super-horizontality. Hence we obtain an action of $\gc$ on the set of those
harmonic maps $M\longrightarrow G/K$ which are of the above form. This is
precisely the action described in Examples 3.5 and 3.6, since in those cases
it turns out that $\lieg_i=0$ for $\vert i\vert>2$, hence (for holomorphic
maps) the concepts of horizontality and super-horizontality coincide. (This
is also, trivially, the action of Example 3.4, where $K=H$.)

Before leaving these examples, we make some brief comments on further
generalizations. It is possible to weaken the hypothesis of
super-horizontality in $(\dagger)$. Indeed, in \cite{BR}, it is shown that
holomorphicity and horizontality, or the even weaker condition of
\lq\lq$J_2$-holomorphicity\rq\rq,  implies that $\varphi$ is harmonic. In
the case $M=S^2$, one then has a converse to $(\dagger)$, namely that any
harmonic map $\varphi:S^2\longrightarrow G/K$ is of the form
$\varphi=\pi\circ\Phi$ for some $J_2$-holomorphic map
$\Phi:S^2\longrightarrow G/H$, for a suitable twistor fibration
$\pi:G/H\longrightarrow G/K$.
These generalizations are not so useful from
the point of view of the action of $\gc$, because neither holomorphicity and
horizontality nor $J_2$-holomorphicity are preserved by this action in
general. On the other hand, there is a natural filtration of $T_{1,0}G/H$ by
the holomorphic subbundles
$T^{(\ell)}=
\gc\times_{G_P}(\bigoplus_{i\le\ell}\lieg_i)/\liegp$.
Let us say that a holomorphic map
$\Phi:M\longrightarrow G/H$ is {\it $\ell$-holomorphic} if it is tangential
to
$T^{(\ell)}$.
Thus, a $1$-holomorphic map is a holomorphic super-horizontal
map; an $\infty$-holomorphic map is simply a holomorphic map. Clearly the
action of $\gc$ preserves $\ell$-holomorphicity. However, the relevance of
this remark depends on the answer to the question: what is the geometrical
significance of the maps $\varphi=\pi\circ\Phi$, where $\Phi$ is
$\ell$-holomorphic?

Finally, we shall explain why the actions in the above examples are special
cases of the natural action of $\flgc$ on extended solutions.
Because of the previous discussion, it suffices to do this for the action of
$\gc$ on $1$-holomorphic maps $\Phi:M\longrightarrow G/H$, where
$G/H=\Ad(G)P$. First, let us define a loop $\ga_P\in\blg$ by
$\ga_P(\la)=\exp\,2\pi tP$, where $\la=e^{2\pi it}$. Then $G/H$ may be
realized as a submanifold of $\blg$, namely as the orbit of $\ga_P$ under
conjugation by $G$. The associated symmetric space $G/K$ may be realized
as a submanifold of $G$, namely as the conjugacy class of $\exp\,\pi P$.
Thus, the twistor fibration $\pi:G/H\longrightarrow G/K$ is just a
restriction of the map $\pi:\blg\longrightarrow G$ (evaluation at $-1$):
$$\CD
G/H   @>>>   \blg   \\
@VVV   @VVV   \\
G/K   @>>>   G
\endCD
$$
Recall that we have the identifications
$T^{\C}_PG/H=\bigoplus_{i\ne 0}\lieg_i$,
and
$T^{\C}_{\ga_P}\blg\cong
T^{\C}_e\blg\cong\bigoplus_{\ell\ne 0}(\la^\ell-1)\liegc$.

\proclaim{Lemma 3.7}
The derivative at $P$ of the embedding
$G/H\longrightarrow\blg$ identifies $\lieg_\ell$ with
\newline $(\la^{-\ell}-1)\lieg_\ell$.
\endproclaim

\demo{Proof} Let $U\in\lieg_\ell$. This corresponds to the initial tangent
vector to the curve $\Ad(\exp\,sU)P$ through $P$ in $G/H=\Ad(G)P$, i.e. to
the curve $(\exp\,sU)\ga_P(\exp\,sU)^{-1}$ through $\ga_P$ in $\blg$. By
left translation we obtain the curve
$\ga_P^{-1}(\exp\,sU)\ga_P(\exp\,sU)^{-1}$ through $e$ in $\blg$.
Now,
$$\align
\ga_P^{-1}(\exp\,sU)\ga_P(\exp\,sU)^{-1}&=
\exp\,\Ad[\exp(-2\pi tP)]sU\,(\exp\,sU)^{-1}\\
&=\exp\,(e^{-2\pi t\ad P}sU)\exp(-sU)\\
&=\exp(e^{-2\pi ti\ell}sU)\exp(-sU)\\
&=\exp(s(\la^{-\ell}-1)U).
\endalign
$$
The initial tangent vector of this curve is $(\la^{-\ell}-1)U$.\qed
\enddemo

In particular, the super-horizontal distribution of $T_{1,0}G/H$ maps into
the subbundle of $T_{1,0}\blg$ defined by $(\la^{-1}-1)\liegc$, so we obtain:

\proclaim{Proposition 3.8}
Via the embedding $G/H\longrightarrow \blg$, a
holomorphic super-horizontal map into $G/H$ goes to an extended solution
into $\blg$. Moreover, the action of $\gc$ on $G/H$ corresponds to the
action of the subgroup $\gc$ of $\flgc$ on $\blg$.\qed\endproclaim

More generally, the concept of $\ell$-holomorphicity for a map
$\Phi:M\longrightarrow G/H$ may be interpreted in terms of the
corresponding map $M\longrightarrow \blg$. Let us say that a holomorphic
map $M\longrightarrow \blg$ is {\it $\ell$-holomorphic} if it is tangential
to
the (holomorphic) subbundle $\Cal H^{(\ell)}$ of $T_{1,0}\blg$ defined by
$\bigoplus_{1\le i\le\ell}(\la^{-i}-1)\liegc$.
Thus, $\ell$-holomorphic maps
interpolate between extended solutions ($\ell=1$) and general holomorphic
maps ($\ell=\infty$).  By Lemma 3.7, $\ell$-holomorphic maps into $G/H$ go
(via
the embedding $G/H\longrightarrow\blg$) to $\ell$-holomorphic maps into
$\blg$. If $\Phi$ is $\ell$-holomorphic, and $\ga\in\flgc$, then
$\ga^\natural\Phi$ is clearly also $\ell$-holomorphic. As in the finite
dimensional case, however, the geometrical significance of maps
$\varphi=\pi\circ\Phi:M\longrightarrow G$, where $\Phi$ is $\ell$-
holomorphic,
is not clear.

In contrast to the actions described in \S 1 and \S 2, the natural
action is very easy to work with. In particular, it has the advantage
that it is always well defined (so we have an action, rather than a
pseudo-action). In the next section we shall give some elementary
properties of this action.

$${}$$

\heading \S 4. Properties of the natural action
\endheading

In this section we shall always take $M$ to be a compact Riemann surface
and $G=U_n$.

The version of the extended solution equations used in the last
section is due to Segal (see \cite{Se}), who used it to give a new proof of
the factorization theorem of \cite{Uh} for harmonic maps
$S^2\longrightarrow U_n$,
and of the classification theorem (see \cite{EL}) for harmonic maps
$S^2\longrightarrow \CP^n$. We shall review Segal's approach here, before
discussing further properties of the natural action. The main
technical result is the following version of Theorem 1.2:

\proclaim{Theorem 4.1 (\cite{Se})} Let
$\Phi: M\longrightarrow\blu$ be an
extended solution. Then there exists a loop $\ga\in\blu$ and a
non-negative integer $m$ such that the map $W=\ga\Phi H_+$
satisfies \newline
(i) $\la^mH_+\subseteq W(z)\subseteq H_+$, for all $z\in M$, and
(ii) $\Span\{W(z)\ \vert \ z\in M\}=H_+$.
Moreover, $m\le n-1$.\qed
\endproclaim

\noindent The extended solution $\ga\Phi$ is said to be {\it
normalized}. For example, let $f: M\longrightarrow\grkcn$ be a holomorphic
map,
so that $\varphi=\pi_f-\pi^\perp_f: M\longrightarrow U_n$ is a harmonic
map (as in
Example 3.4). Then the corresponding extended solution
$\Phi=\pi_f+\la\pi^\perp_f$ is normalized if and only if $f$ is \ll
full\rrr, i.e. $\Span\{f(z)\ \vert\ z\in M\}=\C^n$. It can be shown that
condition (ii) of Theorem 4.1 is equivalent to the Uhlenbeck normalization
(condition (ii) of Theorem 1.2).

Let us assume that $W$ corresponds to a normalized extended
solution, as in the theorem. Then there is a {\it canonical flag}
associated to $W$, namely
$$
\la^mH_+\subseteq W=W_{(m)}\subseteq W_{(m-1)}\subseteq\dots\subseteq
W_{(0)}=H_+
$$
where $W_{(i)}:M\longrightarrow Gr_\infty^{(n)}$ is the holomorphic map
defined by $W_{(i)}=\la^{-(m-i)}W\cap H_+$. Strictly speaking, this
formula defines a holomorphic map with a finite number of
removable singularities, but we shall use the notation $W_{(i)}$ to
mean the map obtained by removing these. The canonical flag
satisfies the following conditions:
$$\gather
\la W_{(i)}\subseteq W_{(i+1)}\tag 0\\
\ddzb \sec W_{(i)}\subseteq\sec W_{(i)}\tag 1\\
\ddz \sec W_{(i)}\subseteq \sec W_{(i-1)}.\tag 2
\endgather
$$
In fact, these equations are {\it equivalent} to the extended solution
equations for $\Phi$, as the holomorphicity condition for $W$ is
given by (1), and the horizontality condition for $W$ follows from
(2) and the definition of $W_{m-1}$. It is an immediate consequence
that each map $W_{(i)}$ satisfies the extended solution equations.
Hence, by the Grassmannian model, we have $W_{(i)}=\Phi_{(i)}H_+$ for
some extended solution $\Phi_{(i)}$.

 From Example 3.5, we see that condition (2) can be interpreted as
saying that the (holomorphic) map $(W_{(i)},W_{(i-1)})$ is horizontal
with respect to the map $(E_{(i)},E_{(i-1)})\longmapsto E_{(i)}^\perp\cap
E_{(i-1)}$. This is why the equation $\ddz\sec
W\subseteq\sec\la^{-1}W$ is called the horizontality condition. Each map
$W_{(i)}^\perp\cap W_{(i-1)}$ is a harmonic map into a
Grassmannian.

 From condition (0) we have
$\la W_{(i-1)}\subseteq W_{(i)}\subseteq W_{(i-1)}$,
so we can derive some further information.
The map $W_{(i-1)}/\la W_{(i-1)}$
defines a holomorphic vector bundle on $M$, and
multiplication by $\Phi_{(i-1)}^{-1}$ defines a smooth isomorphism
of this bundle with the trivial bundle
$M\times H_+/\la H_+\cong M\times\C^n$.
Through this isomorphism, the map
$W_{(i)}/\la W_{(i-1)}$ corresponds to a map $\Psi_i$,
and we have
$\Phi_{(i)}=\Phi_{(i-1)}\Psi_i$. Each map $\Psi_i$ is necessarily of
the form $\pi_{f_i}+\la\pi_{f_i}^\perp$, where $f_i$ is a map from
$M$ to a Grassmannian. By construction, $f_i$ is holomorphic with
respect to a complex structure which is obtained by \ll twisting\rr
the standard complex structure by $\Phi_{(i-1)}$. Hence we have the
factorization theorem: $\Phi$ can be written as
$\Phi=\Psi_1\dots\Psi_m$, where $\Psi_i=\pi_{f_i}+\la\pi_{f_i}^\perp$,
and each sub-product $\Psi_1\dots\Psi_i$ is an extended solution.

This completes our review of \cite{Se}, to which the reader is referred
for further details.
As for the generalization to a pluriharmonic maps,
we can show that Theorem 4.1 holds also for a compact complex manifold
$M$.
Moreover, the above argument for the canonical flag and  the factorization
also
works for the higher dimensional case, if we consider meromorpic maps and
coherent sheaves instead of holomorphic maps and holomorphic vector
bundles
(cf. \cite{OV}).

The finiteness properties of extended solutions described above may
be expressed in terms of a filtration of the \ll algebraic loop
group\rr by certain finite dimensional varieties. The algebraic loop
group is defined as follows:

\noindent{\it Definition.} $\ablu=\{\ga\in\blu\ \vert\ \ga(\la)\
\text{is polynomial in}\ \la,\la^{-1}\}.$

\noindent A Grassmannian model for $\ablu$ may be deduced from
that of $Gr_\infty^{(n)}$:

\proclaim{Proposition 4.2 (\cite{PS})} The image of $\ablu$, under the map
$\blu\longrightarrow Gr_\infty^{(n)}$, is the subspace  $Gr_{\alg}^{(n)}$ of
$Gr_{\infty}^{(n)}$ consisting of linear subspaces $W$ which satisfy
$$
\la^kH_+\subseteq W\subseteq\la^{-k}H_+
\hbox{ for some } k.
$$

\noindent Moreover, if $\ga\in\ablu$ and $W=\ga H_+$, then for such
minimal $k$
we have
$\deg(\det\ga)=\frac12(\dim\,\la^{-k}H_+/W-\dim\,W/\la^kH_+)$.\qed
\endproclaim

\noindent If we define
$$
\alglnc=\{\ga\in\lglnc\ \vert\ \ga(\la),\ga(\la)^{-1}\ \text{are
polynomial in}\ \la,\la^{-1}\},
$$
then we obtain the identification
$$
\ablu\cong\frac{\alglnc}{\La^+_{\alg}\glnc},
$$
where $\alu,\La^+_{\alg}\glnc$ are defined in the obvious way. This is
analogous to the identification $\Omega U_n\cong \Lambda
GL_n(\C)/\Lambda_+GL_n(\C)$ described in the last section. However,
in the case of algebraic loops, one can replace
$\Lambda_{\alg}GL_n(\C)$ by an
even larger group, the semi-direct product
$\CS\ltimes\Lambda_{\alg}GL_n(\C)$, where the action of $\CS$ on
$\Lambda_{\alg}GL_n(\C)$ is given by
\ll re-scaling\rrr, i.e. $(v\cdot\ga)(\la)=\ga(v^{-1}\la)$ for all $v\in\CS$,
$\ga\in\Lambda_{\alg}GL_n(\C)$. The group $\CS$ also acts on
$\Omega_{\alg}U_n$, by
$$
v\nat\ga=[v\cdot\ga]
$$
where square brackets (as usual) denote cosets in $\Lambda_{\alg}GL_n(\C)/
\Lambda^+_{\alg}GL_n(\C)\cong\Omega_{\alg}U_n$. We use the \ll natural\rr
notation for this action, because the formula
$$
(v,\ga)\nat\de = \ga\nat v\nat\de
$$
defines an action of $\CS\ltimes\Lambda_{\alg}GL_n(\C)$ on
$\Omega_{\alg}U_n$, which extends the natural action of
$\Lambda_{\alg}GL_n(\C)$ on $\Omega_{\alg}U_n$. Thus we obtain finally
the identifications
$$
\ablu\cong\frac{\alglnc}{\La^+_{\alg}\glnc}\cong
\frac{\CS\ltimes\alglnc}{\CS\ltimes\La^+_{\alg}\glnc}.
$$

Mitchell (\cite{Mi}) introduced the following subspaces of
$Gr_{\alg}^{(n)}$ (see also \S 1 of \cite{Se}):

\noindent{\it Definition.} $F_{n,k}=\{W\subseteq H^{(n)}\ \vert\
\la^kH_+\subseteq W\subseteq H_+,\ \la W\subseteq W, \dim\,H_+/W=k\}$.

 \noindent It can be shown that $F_{n,k}$ is a connected complex algebraic
subvariety of the
Grassmannian $Gr_{kn-k}(\C^{kn})$. Explicitly, if we make the
identification
$\C^{kn}\cong H_+/\la^kH_+ =\allowmathbreak
\Span\{[\la^{i}e_j]\ \vert\ 0\le i\le k-1, 1\le j\le n\}$, then
$$
F_{n,k}\cong\{E\in Gr_{kn-k}(\C^{kn})\ \vert\ NE\subseteq E\},
$$
where $N$ is the nilpotent operator on $\C^{kn}$ given by the
multiplication by $\la$. The space $F_{n,k}$ is preserved by the
action of $\La_+\glnc$, since (by definition) this group fixes $H_+$.
The action of $\La_+\glnc$ on $F_{n,k}$ collapses to the action of
the finite dimensional group
$$
G_{n,k}=\{X\in GL_{kn}(\C)\ \vert\ XN=NX\}.
$$
Indeed, the action of $\La_+\glnc$ factors through the homomorphism
$\La_+\glnc\longrightarrow GL_n(\C[\la]/(\la^k))$ defined by
$\sum_{i\ge0}\la^iA_i\longmapsto\sum_{i=0}^{k-1}\la^iA_i$, and we have
$GL_n(\C[\la]/(\la^k))\cong G_{n,k}$. This is a complex Lie group of
dimension $kn^2$. The action of $\CS$ also preserves $F_{n,k}$, for the
action of an element $u\in\CS$ induces the linear transformation
$T_u[\la^ie_j]=[u^i\la^ie_j]$, and so $T_uN=uNT_u$.

In these terms, we see that a normalized extended solution is a \ll
horizontal\rr holomorphic map $\Phi:M\longrightarrow F_{n,k}$, where the
integer $k=\deg\,\det\Phi(z)$ represents the connected component
of $\blu$ which contains the image of $\Phi$. The minimal uniton
number $m$ satisfies the conditions $m\le n-1$, $m\le k$. It is
known (see \cite{Mi}) that $H_2(F_{n,k};\Z)\cong\Z$, so $\Phi$ has a
topological \ll degree\rr $d$, which (with appropriate choice of
orientations) is a non-negative integer. The
geometrical significance of $d$ is that it represents the energy of
the corresponding harmonic map $\varphi:M\longrightarrow U_n$ (see
\cite{EL},\cite{Va},\cite{OV}). From the discussion
above we have:

\proclaim{Proposition 4.3} The natural action of $\La_+\glnc$ on
normalized extended solutions $\Phi$ preserves
\roster
\item the connected component $k$ of $\blu$ containing the image
of $\Phi$,
\item the minimal uniton number $m$ of $\Phi$, and
\item the degree $d$ of $\Phi$ (i.e. the energy of the corresponding
harmonic map).
\endroster
Moreover, for a fixed choice of $k$, the action of $\La_+\glnc$ on
normalized extended solutions collapses to the action of the finite
dimensional (complex) Lie group $G_{n,k}$.
\qed
\endproclaim

\noindent In this proposition, $\La_+\glnc$ could be replaced by the group
$\CS\ltimes\La^+_{\alg}\glnc$, and $G_{n,k}$ by $\CS\ltimes G_{n,k}$;
we leave the verification of this to the
reader. The existence of an action of $\CS$ on
extended solutions was first noticed by Terng (see \S 7 of \cite{Uh}).

$${}$$

\heading
\S{5}. Relation between the Uhlenbeck pseudo-action
and the natural action
\endheading

In this section we shall show that
the Uhlenbeck pseudo-action discussed in \S 2 and the
natural action defined in \S 3 coincide
on harmonic maps of finite uniton number.
We begin by considering a special case.

For any $\ep$ with $0<\ep<1$,
we have an injective homomorphism as real Lie groups
$$
\Lambda_{+}\glnc\longrightarrow
\Lambda_{I,{\R}}\glnc\subseteq{\Cal G}_{\bold R},
\quad\ga\longmapsto\hat\ga
$$
defined by
$$
\hat{\ga}(\la)
=\cases
\ga(\la)
&\text{for $\vert\la\vert\leq\ep$},\\
\ga(\bar{\la}^{-1})^{-1\ast}
&\text{for $\vert\la\vert\geq{1/\ep}$}\\
\endcases
$$
for
$\ga\in\Lambda_{+}GL_n({\C})$.

\proclaim {Theorem 5.1}
If
$\ga\in\Lambda_{+}GL_n({\C})$
and
$\delta\in{\Cal X}_{k,{\R}}\subseteq{\Omega}U_n$
for
$0\leq{k}\leq{\infty}$,
then
$\hat{\ga}^{\sharp}\delta
\in{\Cal X}_{k,{\R}}$ is well-defined and
$$
\ga^{\natural}\delta=\hat{\ga}^{\sharp}\delta.
$$
\endproclaim

\demo {Proof} By the decomposition
$\Lambda GL_n({\C})\cong\Omega{U_n}\cdot
\Lambda_{+}GL_n({\C})$,
we have
$
\ga\delta=(\ga\delta)_u(\ga\delta)_{+}$,
where
$(\ga\delta)_u\in\Omega{U_n},
(\ga\delta)_{+}\in\Lambda_{+}GL_n({\C}).
$
Note that
$(\ga\delta)_u(\la)=
\ga(\la)\delta(\la)
(\ga\delta)_{+}^{-1}(\la)$
extends holomorphically to
$\{\la\in{\C} \ \vert\  0<\vert\la\vert<1\}$.
Define
$$
(\hat\ga\delta)_I(\la)
=
\cases
(\ga\delta)_{+}(\la)
&\text{for $\vert\la\vert\leq\ep$},\\
\{(\ga\delta)_{+}(\bar\la^{-1})\}^{-1\ast}
&\text{for $\vert\la\vert\geq{1/\ep}$},\\
\endcases
$$
namely,
$(\hat\ga\delta)_I=
(\ga\delta)\,\hat{}_{+}
\in\Lambda_{I,{\R}}\glnc$.
Define
$$
(\hat\ga\delta)_E(\la)
=
\cases
(\ga\delta)_{u}(\la)
&\text{for $0<\vert\la\vert\leq 1$},\\
\{(\ga\delta)_{u}(\bar\la^{-1})\}^{-1\ast}
&\text{for $\quad 1\leq\vert\la\vert<\infty$}.
\endcases
$$
By Painlev\'e's Theorem we have
$(\hat\ga\delta)_E\in
\Lambda_{E,{\R}}\glnc$,
and moreover
$(\hat\ga\delta)_E\in{\Cal X}_{k,{\R}}$,
because
$(\ga\delta)_u=
\ga\delta(\ga\delta)_{+}^{-1},
\delta\in{\Cal X}_{k,{\R}}$.

For $0<\vert\la\vert\leq{\ep}$, we have
$$
\align
\quad(\hat\ga\delta(\hat\ga\delta)_I^{-1})
(\la)
&=\ga(\la)\delta(\la)
(\ga\delta)_{+}^{-1}(\la)\\
&=(\ga\delta)_u(\la)=(\hat\ga\delta)_E(\la).
\endalign
$$
For $1/{\ep}\leq\vert\la\vert<\infty$, we have
$$
\align
\quad(\hat\ga\delta(\hat\ga\delta)_I^{-1})
(\la)
&=\ga(\bar\la^{-1})^{-1\ast}
\delta(\bar\la^{-1})^{-1\ast}
\{(\ga\delta)_{+}(\bar\la^{-1})\}^{\ast}\\
&=\{\ga(\bar\la^{-1})
\delta(\bar\la^{-1})
(\ga\delta)_{+}(\bar\la^{-1})\}^{-1\ast}\\
&=\{(\ga\delta)_u(\bar\la^{-1})\}^{-1\ast}=(\hat\ga\delta)_E(\la).
\endalign
$$
Hence
$\hat\ga\delta(\hat\ga\delta)_I^{-1}
=(\hat\ga\delta)_E=\hat\ga^{\sharp}\delta$.
Thus we obtain
$\hat\ga^{\sharp}\delta
=(\ga\delta)_u
=\ga^{\natural}\delta$.
\qed
\enddemo

\proclaim{Corollary 5.2}
If $\ga\in\Lambda_{+}GL_n(\C)$
and $\Phi : M\longrightarrow \Omega{U_n}$
is an extended solution such that $\Phi_{\la}$
is holomorphic in $\la\in{\C}^{\ast}$, then we
have
$$
\ga^{\natural}\Phi=\hat{\ga}^{\sharp}\Phi.
$$
\endproclaim

\demo {Proof}
By assumption the image of $\Phi$ is contained in
${\Cal X}_{\R}={\Cal X}_{\infty,\R}$.
Hence the corollary follows from Theorem 5.1.
\qed
\enddemo

In \S 2, we saw that the Uhlenbeck pseudo-action of ${\Cal G}_{\R}$ on
${\Cal X}_{k,{\R}}$ collapses to the pseudo-action of the finite dimensional
Lie group
${\Cal G}_{\R}/{\Cal G}_{k,\R}\cong{\Cal A}_{\R}/{\Cal A}_{k,\R}$, and
in \S4  that the natural action of $\La_+\glnc$ on $F_{n,k}$ collapses to the
action of the Lie group $G_{n,k}$.
Evidently, we have ${\Cal G}_{\R}/{\Cal G}_{k,\R}\cong G_{n,k}$
as real Lie groups.
{}From Theorem 5.1, and by using the same argument as was
used at the end of \S2,
we see that the pseudo-action of ${\Cal G}_{\R}$ (or ${\Cal G}_{\R}/{\Cal
G}_{k,\R}$) on extended solutions with finite uniton number is an action,
and coincides with the action of  $\La_+\glnc$ (or $G_{n,k}$). Hence:

\proclaim{Corollary 5.3}
The Uhlenbeck pseudo-action of ${\Cal G}_{\R}$ on
extended solutions (or harmonic maps) with finite uniton number
coincides with the natural action of $\La_+\glnc$.\qed
\endproclaim

$${}$$

\heading \S 6. Deformations of harmonic maps
\endheading

Let $\{g_t\}$ be a curve in $\flgc$, i.e. a continuous map
$t\longmapsto g_t$ from an open interval of $\R$ to $\flgc$,
with $g_0=e$. Let $\Phi:M\longrightarrow\blg$ be an extended solution.
Then the formula
$$
\Phi_t=g\nat_t\Phi
$$
defines a continuous family of extended solutions passing through
$\Phi$ (a \ll deformation\rr of $\Phi$). For example, we can take
$\{g_t\}$ to be a one parameter subgroup $\{\exp\,t\be\}$, for
$\be\in\fl\frak g^{\C}$. The same observation applies to a curve in
$\CS\ltimes\Lambda_{\alg}\glnc$, providing that the extended solution
$\Phi$ takes values in $\Omega_{\alg}G$.

Now, it may happen that $\lim_{t\to\infty}\Phi_t$ exists, even if
$\lim_{t\to\infty}g_t$ does not exist, and in this case we obtain an
extended solution $\Phi_\infty=\lim_{t\to\infty}\Phi_t$ which is
{\it not}, a priori, of the form $g\nat\Phi$. Some examples of this
\ll completion\rr process were studied in \cite{BG} for the case of the
Uhlenbeck action $\sharp$. By using the action $\natural$, however,
we can obtain more detailed information. The reason for this is that,
for certain $\be$, the curve $\ga\longmapsto(\exp\,t\be)\nat\ga$ has a
simple geometrical interpretation: it is a flow line of the gradient
vector field of a natural Morse-Bott function on $\blg$.

The basic example of a Morse-Bott function on $\blg$ is the \lq\lq perturbed
energy functional\rq\rq \
\newline $E+cK^Q$, where $E$ is the energy functional
$$
E(\ga)=\frac12\int_{S^1}\ \vert\vert\ga^{-1}\ga^\prime\vert\vert^2,
$$
and $K^Q$ is the momentum functional
$$
K^Q(\ga)=\int_{S^1}\ \langle\langle\ga^{-1}\ga^\prime,Q\rangle\rangle,
$$
for some fixed $Q\in\lieg$, and where $c$ is a non-zero constant. The
critical points of $E+cK^Q$ are simply the homomorphisms
$S^1\longrightarrow C(T_Q)$, where $C(T_Q)$ is the centralizer in $G$ of
the torus $T_Q$ generated by $Q$. It is classical that this is a Morse-Bott
function. The flow of $-\nabla E$ with respect to
the K\"ahler metric is given by the re-scaling action of the one parameter
semi-group
$\{e^{-t}\ \vert\ t\ge0\}$, and the flow of $-\nabla K^Q$ is given by the
(natural) action of
$\{\exp\,itQ\}$. Hence the flow of $E+cK^Q$ is given by the action of
$\{\exp\,it(i,cQ)\ \vert\ t\ge0\}$ (which is contained in $\CS\times\gc$,
and hence in $\CS\ltimes\La_{\alg}G^{\C}$).

As a first application, let us consider the case where $Q$ is a regular point
of $\lieg$. Since $Q$ generates (by definition) a maximal torus $T$, which is
equal to its own centralizer, the critical points are the homomorphisms
$S^1\longrightarrow T$; in particular, they are isolated. The stable manifold
of a critical point is a cell in $\blg$ of finite codimension, the so called
Birkhoff cell (see \cite{PS}). If $\Phi:M\longrightarrow \blg$ is a
holomorphic map, then $\Phi(z)$ must lie in a single Birkhoff manifold for
all but a finite number of points $z\in M$, so we obtain:

 \proclaim{Proposition 6.1} Let $\Phi:M\longrightarrow\Omega_{\alg}G$ be
an extended solution. Then there exists a curve $\{g_t\}$ in
$\CS\ltimes\La_{\alg} \gc$ such
that
$$
\Phi_\infty(z)=\lim_{t\to\infty}g_t\nat\Phi(z)
$$
defines a constant (extended solution) $\Phi_\infty:M\setminus
S\longrightarrow\Omega G$, where the set $S$ consists of a finite number
of
removable singularities of $\Phi_\infty$.\qed
\endproclaim

\noindent This is an example of the \ll bubbling off\rr phenomenon for
harmonic maps (\cite{SU}). Proposition 6.1 answers positively the question
posed at the end of \S 7 of \cite{BG},
namely whether any extended solution can be reduced to a constant map by
applying the \ll modified completion\rr procedure. However, it is perhaps of
more interest to find deformations where the singularities do not occur, and
this we shall do next.

 For our second application, we shall consider the function $K^Q$. The set of
critical points is $\Omega C(T_Q)$, which is infinite dimensional. However,
for the application to extended solutions, we are primarily interested in the
restriction of $K^Q$ to the finite dimensional subvariety $F_{n,k}$ (with
$G=U_n$). Let us now consider the flow of $-\nabla K^Q$, which is given by
the natural action of $\{\exp\,itQ\}$ on $\blu$. We may consider
$\{\exp\,itQ\}$ to be a one parameter subgroup of $\La_+\glnc$, so it
preserves $F_{n,k}$. (Indeed, $\{\exp\,itQ\}$ is a one parameter subgroup of
$\glnc=G_{n,1}\subseteq G_{n,k}$, in the notation of \S4.) We shall use this
flow, with a suitable choice of $Q$, to prove:

\proclaim{Theorem 6.2} Let
$\Phi=\sum_{\al=0}^mT_\al\la^\al:M\longrightarrow F_{n,k}$ be a
normalized extended solution. If $\rank\,T_0(z)\ge2$ for all $z\in M$,
then $\Phi$ can be deformed continuously through extended solutions to an
extended solution $\Psi:M\longrightarrow\Omega U_{n-1}$.
\endproclaim

\demo{Proof} Let $Q=i\pi_L$ where $\pi_L:\C^n\longrightarrow\C^n$
denotes orthogonal projection onto a complex line $L$ in $\C^n$. The
homomorphism $\glnc\lra G_{n,1}\lra G_{n,k}\sub GL_{kn}(\C)$ will be
denoted $X\longmapsto X^\prime$. Thus, if $\C^{kn}$ is identified with
$H_+/\la^kH_+$ as usual, we have $X^\prime(\la^iv)=\la^iXv$ for any
$v\in\C^n$. Observe that $(\pi_L)^\prime=\pi_{L^\prime}$, where $L^\prime$
is the $k$-plane $L\oplus\la L\oplus\dots\oplus\la^{k-1}L$.

Consider the flow on the Grassmannian $\gr$ which is given by the action of
the one parameter subgroup $\{(\exp\,itQ)^\prime\}$
($=\{\exp\,itQ^\prime\}$) of $GL_{kn}(\C)$. It is well known that this is the
downwards gradient flow of a Morse-Bott function on $\gr$, such that

\noindent (1) the set of absolute minima is $G^L=\{W\ \vert\ L^\prime\sub
W\}$, and

\noindent (2) the stable manifold of $G^L$ (i.e. the union of the flow lines
which terminate on $G^L$) is $S^L=\{W\ \vert\ W^\perp\cap
L^\prime=\{0\}\}$.

\noindent (These assertions represent a mild generalization of the standard
Schubert cell decomposition of a Grassmannian. They are explained in more
detail in the Appendix.)

Observe that $G^L\cap F_{n,k}=F_{n-1,k}$ if we take $L=\Span\{e_n\}$. Thus,
if the image of the extended solution $\Phi$ is contained entirely in
$S^L\cap F_{n,k}$, the formula $\Phi_t=(\exp\,itQ^\prime)\nat\Phi$ gives a
continuous deformation of $\Phi$ into $F_{n-1,k}$. To prove the theorem,
therefore, it suffices to show that any extended solution satisfying the
hypotheses lies in $S^L\cap F_{n,k}$, for {\it some} line $L$.

Let $\Phi$ be an extended solution satisfying the hypotheses. Let
$$
Y^\Phi=\{L\ \vert\ \Phi(z)\notin S^L\ \text{for some}\ z\in M\}.
$$
Thus, $Y^\Phi$ is the set of \lq\lq bad\rq\rq\ lines in $\C^n$. We shall show
that $\dim_{\C}Y^\Phi<n-1$, which implies immediately that not all lines
are \lq\lq bad\rq\rq.

To do this, note that
$$\align
\Phi(z)\notin S^L &\iff \Phi(z)^\perp\cap L^\prime\ne\{0\}\\
&\iff \Phi(z)^\perp\cap L\ne\{0\}\\
&\iff\Phi(z)\sub L^\perp
\endalign
$$
(the middle step follows from the fact that both $\Phi(z)^\perp$ and
$L^\prime$ are preserved by the adjoint of multiplication by $\la$, i.e. by
the linear transformation $\la^\ast$ of $H_+/\la^kH_+$ given by
$\la^\ast(\la^ie_j)=\la^{i-1}e_j$, $1\le i\le k-1$, and $\la^\ast(e_j)=0$).
Let $X=\{(L,W)\in\CP^{n-1}\times F_{n,k}\ \vert\ W\sub L^\perp\}$. Let
$p_1:X\lra\CP^{n-1}$, $p_2:X\lra F_{n,k}$ be the projection maps. Then we
have $Y^\Phi=p_1(p_2^{-1}(\Phi(M)))$, so
$\dim_{\C}Y^\Phi\le\dim_{\C}p_2^{-1}(\Phi(M))$. We claim that
$\dim_{\C}p_2^{-1}(\Phi(z))\le n-3$ for all $z\in M$. Since $\dim_{\C}M=1$,
we may then conclude that $\dim_{\C}Y^\Phi<n-1$, as required.
{}From the expression $\Phi=\sum_{\al=0}^mT_\al\la^\al$ we see that
$$
p_2^{-1}(\Phi(z))=\{L\ \vert\ \Phi(z)\sub L^\perp\}=\bold
P(\Ker\,T_0^\ast(z)),
$$
so the claim follows from the hypothesis.
\qed\enddemo

It is appropriate at this point to make some comments on the use of Morse
theory in the proof of Theorem 6.2. The fact that $F_{n,k}$ is in general a
singular variety (to which ordinary Morse theory does not apply) is
irrelevant for our purposes, as we are concerned only with the given flow.
However, to study this flow in practice, it is useful to regard it as the
restriction of a flow on the Grassmannian $Gr_{kn-k}(\C^{kn})$, where it is
indeed the downwards gradient flow of a Morse-Bott function. This type of
Morse-Bott function is well understood: it is an example of a \lq\lq height
function\rq\rq\  on an orbit of the adjoint representation of a compact Lie
group. In the Appendix to this paper, we summarize the basic facts
concerning such height functions. Briefly, the situation is as follows.
Consider a
finite dimensional generalized flag manifold of $G$, i.e. an orbit
$\Ad(G)P$ of a point $P$ of $\lieg$ under the adjoint representation.
Let $Q$ be any
element of $\frak g$. Then one may define the height function
$h^Q:\Ad(G)P\longrightarrow\R$ by $h^Q(X)=
\langle\langle X,Q\rangle\rangle$. This is a Morse-Bott function
and its non-degenerate critical manifolds can be described
explicitly in Lie theoretic terms. Let $\nabla h^Q$ be the gradient of $h^Q$
with respect to the natural K\"ahler metric on $\Ad(G)P$. Then the flow line
of $-\nabla h^Q$ which passes through a point $X$ of $\Ad(G)P$ is given by
$t\longmapsto(\exp\,itQ)\nat X$.

This can be used to obtain results analogous to Theorem 6.2 for harmonic
maps $M\longrightarrow G/K$, for various inner symmetric spaces $G/K$,
because the total space of the corresponding twistor fibration is a
generalized flag manifold. Although this is simply a special case of the
discussion above, it is instructive to give a direct argument (avoiding the
paraphernalia of extended solutions), and this we shall do for each of the
three examples considered in \S3. This will, incidentally, provide some
examples
of extended solutions $\Phi$ which satisfy the hypotheses of Theorem 6.2.

\noindent{\bf Example 6.3 (cf. Example 3.4).} Let
$\text{Hol}_d(S^2,\grkcn)$ denote the space of holomorphic maps
$\Phi:S^2\longrightarrow\grkcn$ which have degree $d$. It is well known
that
this space is connected. However, we shall give a proof of this fact as an
illustration of the technique introduced above.

We identify $\grkcn$ with the orbit $\Ad(U_n)P$ in $\frak u_n$,
where $P=i\pi_V$ for some $k$-plane $V$. Let $Q=i\pi_1$, where
$\pi_1:\C^n\longrightarrow \C$ denotes orthogonal projection onto the line
spanned by the first standard basis vector. The action of the one
parameter subgroup $\{\exp\,itQ\}$ gives the downwards gradient flow of a
Morse-Bott function $\grkcn\longrightarrow\R$. (See the Appendix.) The
critical points are those
$k$-planes $W\in\grkcn$ for which $[i\pi_1,i\pi_W]=0$, i.e. for
which $\C\subseteq W$ or $W\subseteq\C^\perp$.  Thus there are two
connected critical manifolds:
$$\gather
G^+=\{W\ \vert\ \C\subseteq W\}\cong Gr_{k-1}(\C^{n-1})\\
G^-=\{W\ \vert\ W\subseteq\C^\perp\}\cong Gr_k(\C^{n-1}).
\endgather
$$
The corresponding stable manifolds are:
$$\gather
S^Q(G^+)=G^+\\
S^Q(G^-)=\{W\ \vert\ W\cap\C=\{0\}\}.
\endgather
$$
We claim that the inclusions
$$
\text{Hol}_d(S^2,Gr_k(\C^{n-1}))\cong\text{Hol}_d(S^2,G^-)
\longrightarrow\text{Hol}_d(S^2,S^Q(G^-
))\longrightarrow\text{Hol}_d(S^2,\grkcn)
$$
induce bijections on the sets of connected components. In the case
of the first inclusion, this is so because, if $\Phi(S^2)\subseteq S^Q(G^-)$,
then $\{(\exp\,itQ)\nat\Phi\}_{0\le t\le\infty}$ provides a continuous
deformation of $\Phi$ into $G^-$. For the second inclusion, it is
because $\text{Hol}_d(S^2,S^Q(G^-))$ is obtained from the manifold
$\text{Hol}_d(S^2,\grkcn)$ by removing a closed subvariety of
complex codimension $1$. By induction it follows that
$\text{Hol}_d(S^2,\grkcn)$ has the same number of connected
components as $\text{Hol}_d(S^2,\C P^k)$. However, from the usual
description of holomorphic maps $S^2\longrightarrow \CP^k$ in terms of
polynomials, it follows that this space is connected.

By modifying this argument slightly (see the proof of Theorem 6.5 below), it
can be shown that $\Hol_d(M,\grkcn)$ is connected for any compact Riemann
surface $M$, providing that $d\ge 2g$, where $g$ is the genus of $M$. The last
restriction ensures that $\Hol_d(M,S^2)$ is connected
(see Corollary 1.3.13 of \cite{Na}). In fact, these conditions may be weakened
;
 for
example in \cite{To} it is shown that $\Hol_d(M,S^2)$ is connected when $d\ge
 g$, and it
follows from  \cite{FL} that $\Hol_d(M,S^2)$ is connected for \ll generic \rr
 $M$ when
$d\ge (g+3)/2$.

\noindent{\bf Example 6.4 (cf. Example 3.5).} Let
$\text{Harm}_d(S^2,\C P^n)$ denote the space of harmonic maps
$\varphi:S^2\longrightarrow\C P^n$ which have degree $d$. If
$\Phi:S^2\longrightarrow
F_{r,r+1}(\C^{n+1})$ corresponds to a harmonic map $\varphi$ as in
Example 3.5, and if $\deg\Phi=(\deg\,W_r,\deg\,W_{r+1})=(k,l)$,
then we have
$$
d=l-k,\ \ \ E=l+k
$$
where $E$ denotes the (suitably normalized) energy. If $n>1$, it is
easy to construct examples of harmonic maps $\varphi_1,\varphi_2$ with
$\deg\varphi_1=\deg\varphi_2$ but $E(\varphi_1)\ne E(\varphi_2)$, so the
space
of harmonic maps of fixed degree cannot be connected. However, we
can prove:

\proclaim{Theorem 6.5} (i) The inclusion
$\text{Harm}_d(S^2,\C P^2)\longrightarrow\text{Harm}_d(S^2,\C P^n)$
induces a bijection on the sets of connected components, if $n\ge2$.
(ii) More generally, the same is true if harmonic maps from $S^2$ are
replaced by complex isotropic harmonic maps from any compact Riemann
surface $M$.
\endproclaim

\demo{Proof} It suffices to give the proof of (ii).
Let $\Phi=(W_r,W_{r+1}):M\longrightarrow F_{r,r+1}(\C^{n+1})$ be a
holomorphic horizontal map associated to $\varphi$. We shall use the
method of Example 6.3 to show that $\Phi$ may be deformed into
$F_{r,r+1}(\C^n)$, if $r<n-1$. Hence, by induction, we obtain a map
(also denoted by $\Phi$) whose image lies in $F_{r,r+1}(\C^{r+2})$.
By repeating this argument with
$\Phi^\ast=(W_{r+1}^\perp,W_r^\perp)$, we can similarly deform
$\Phi$ into $\{(E_r,E_{r+1})\in F_{r,r+1}(\C^{r+2})
\ \vert\ \C^{r-1}\subseteq E_r\}$. Thus we obtain a deformation of
$\varphi$
into $P(\C^{r+2}/\C^{r-1})$, and hence (by applying a projective
transformation) into $\C P^2$.

We identify $F_{r,r+1}(\C^{n+1})$ with the orbit
$\Ad(U_{n+1})(i\pi_{V_r}+i\pi_{V_{r+1}})$, where $(V_r,V_{r+1})$ is
a fixed element of $F_{r,r+1}(\C^{n+1})$. Let $\pi_n^\perp$ denote
orthogonal projection onto the line $(\C^n)^\perp$ in $\C^{n+1}$ spanned by
the last standard basis vector, and set $Q=i\pi_n^\perp$.
We shall use the Morse-Bott function on $F_{r,r+1}(\C^{n+1})$ whose
downwards gradient flow is given by the action of $\{\exp\,itQ\}$.
A point
$(E_r,E_{r+1})$ is a critical point
if and only if $[i\pi_n^\perp,i\pi_{E_r}+i\pi_{E_{r+1}}]=0$, i.e. the
line $(\C^n)^\perp$ is contained in $E_r$, $E_r^\perp\cap E_{r+1}$, or
$E_{r+1}^\perp$. The three connected critical manifolds are:
$$\gather
F^+=\{(E_r,E_{r+1})\ \vert\ (\C^n)^\perp\subseteq E_r\}=F_{r-1,r}(\C^n)\\
F^0=\{(E_r,E_{r+1})\ \vert\ (\C^n)^\perp= E_r^\perp\cap
E_{r+1}\}\cong Gr_r(\C^n)\\
F^-=\{(E_r,E_{r+1})\ \vert\ (\C^n)^\perp\subseteq
E_{r+1}^\perp\}=F_{r,r+1}(\C^n).
\endgather
$$
The corresponding stable manifolds are:
$$\gather
S^Q(F^+)=F^+\\
S^Q(F^0)=\{(E_r,E_{r+1})\ \vert\ (\C^n)^\perp\subseteq E_{r+1},\
(\C^n)^\perp\cap E_r=\{0\}\}\\
S^Q(F^-)=\{(E_r,E_{r+1})\ \vert\ (\C^n)^\perp\cap E_{r+1}=\{0\}\}.
\endgather
$$
If $\Phi(S^2)\subseteq S^Q(F^-)$, then $\{(\exp\,itQ)\nat\Phi\}_{0\le
t\le\infty}$ provides a continuous deformation of $\Phi$ into $F^-
=F_{r,r+1}(\C^n)$. So it suffices to show that $\Phi$ can be deformed
into $S^Q(F^-)$. In Example 6.3, the corresponding fact was true for
dimensional reasons, but a different argument is necessary in the
present situation as the space of holomorphic horizontal maps is not
in general a manifold. (The argument we are about to give is also needed in
Example 6.3, in the case of a Riemann surface.)

We claim that there exists some $A\in U_{n+1}$ such that
$A\nat\Phi(M)\subseteq S^Q(F^-)$, i.e.
$AW_{r+1}(z)\not\supseteq(\C^n)^\perp$
for all $z\in M$; from this one can construct the required
deformation, as $U_{n+1}$ is connected. It suffices to find some line
$L$ such that $W_{r+1}(z)\not\supseteq L$ for all $z\in M$. Let
$$
Y^\Phi=\{L\in\C P^n\ \vert\ L
\subseteq W_{r+1}(z)\ \text{for some}\ z\in M\}.
$$
Then our claim is that $Y^\Phi\ne\C P^n$. Let
$X=\{(L,E_r,E_{r+1})\in\C P^n\times F_{r,r+1}(\C^{n+1})\ \vert\
L\subseteq E_{r+1}\}.$
Let $p_1$ and $p_2$ be the projections to $\C P^n$ and
$F_{r,r+1}(\C^{n+1})$.
Then $Y^\Phi=p_1(p_2^{-1}(\Phi(M)))$. We have
$\dim_{\C} Y^\Phi\le
\dim_{\C} p_2^{-1}(\Phi(M))\le r+\dim_{\C}\Phi(M)$
(as
the fibre of $p_2$ is $\C P^r$) $\le r+1$. Hence $Y^\Phi$ cannot be
equal to $\C P^n$ if $r<n-1$. This completes the proof.
\qed\enddemo

\noindent{\it Remark:} We have extended solutions of the form
$\pi_f+\la\pi_f^\perp$ in Example 6.3, and
$(\pi_{f_r}+\la\pi_{f_r}^\perp)(\pi_{f_{r+1}}+\la\pi_{f_{r+1}}^\perp)$ in
Example 6.4. It follows that the deformations used in these examples could
have been obtained by applying Theorem 6.2, because the deformation of
Theorem 6.2 preserves the relevant Grassmannian or flag manifold and the
hypotheses of that theorem are satisfied.

\noindent{\bf Example 6.6 (cf. Example 3.6).} Let
$\text{Harm}_d(S^2,S^n)$ be the space of harmonic maps
$\varphi:S^2\longrightarrow
S^n$ of energy $d$, with a similar definition for
$\text{Harm}_d(S^2,\R P^n)$.

\proclaim{Theorem 6.7} (i) $\text{Harm}_d(S^2,S^n)$ and
$\text{Harm}_d(S^2,\R P^n)$ are connected, if $n\ge 3$.
\newline (ii) More generally, the space of isotropic harmonic maps of energy
$d$ of any compact Riemann surface $M$ into $S^n$ (or $\R P^n$) is
connected, if $n\ge 3$ and if $d\ge 2g$, where $g$ is the genus of $M$.
\endproclaim

\noindent{\it Remark:} This result is elementary if $n=3$. Part (i)
was proved by Loo (\cite{Lo}) and by Verdier (\cite{Ve3}) for $n=4$, and
extended to $n\ge4$ by Kotani
(\cite{Kt}).

\demo{Proof} It suffices to give the proof of (ii).
The result for $S^n$ follows from that for $\R P^n$, as
the natural map $S^n\longrightarrow\R P^n$ induces a non-trivial double
covering
$\text{Harm}^{iso}_d(M,S^n)\longrightarrow\text{Harm}^{iso}_d(M,\R P^n)$,
where  $\text{Harm}^{iso}_d$ denotes isotropic harmonic maps of energy
$d$. By \cite{Ca1},\cite{Ca2}
it suffices to take $n$ even, say $n=2m$, and it suffices to show
that the space $\HH_d(S^2,Z_m)$ of holomorphic horizontal maps
$\Phi:M\longrightarrow Z_m$ of degree $d$ is connected, as the map
$\pi:Z_m\longrightarrow\R P^{2m}$ induces a surjection
$\HH_d(M,Z_m)\longrightarrow\text{Harm}^{iso}_d(M,\R P^{2m})$. (The
degree of
$\Phi$ is equal to the energy of $\varphi=\pi\circ\Phi$, if the energy is
normalized suitably.)

We shall prove that $\HH_d(M,Z_m)$ is connected by induction on
$m$. For $m=1$, the horizontality condition is vacuous, so
$\HH_d(M,Z_m)$ may be identified with the space
$\text{Hol}_d(M,S^2)$.
This is known to be connected if $d\ge 2g$ (see Corollary 1.3.13 of \cite{Na}
and also the comments in Example 6.3),and so the
induction begins.

For the inductive step, we shall identify $Z_m$ with the orbit
$\Ad(SO_{2m+1})(i\pi_V-i\pi_{\bar V})$, where $V$ is a fixed
element of $Z_m$. Let $L$ be an isotropic line in $\C^{2m+1}$, and
set $Q=i\pi_L-i\pi_{\bar L}$. The critical points $W\in Z_m$ of the
Morse-Bott function whose downwards gradient flow is given by the action
of $\{\exp\,itQ\}$
 are given by $[i\pi_W-i\pi_{\bar
W},i\pi_L-i\pi_{\bar L}]=0$, i.e. $W=W_1\oplus W_2\oplus W_3$
with $W_1\subseteq L$, $W_2\subseteq\bar L$, $W_3\subseteq(L\oplus
\bar L)^\perp$. There are two connected critical manifolds, namely:
$$\gather
Z^+=\{W\ \vert\ L\subseteq W\subseteq\bar L^\perp\}\cong Z_{m-1}\\
Z^-=\{W\ \vert\ \bar L\subseteq W\subseteq L^\perp\}\cong Z_{m-1}.
\endgather
$$
The corresponding stable manifolds are
$$\gather
S^Q(Z^+)=Z^+\\
S^Q(Z^-)=\{W\ \vert\ W\cap L=\{0\}\}.
\endgather
$$
The embeddings $I^\pm:Z_{m-1}\longrightarrow Z_m$ defined by the
inclusions of
$Z^\pm$ in $Z_m$ are holomorphic. They also respect the
horizontality condition $\ddz\sec\Phi\perp\sec\bar\Phi$, in the
sense that a map $\Phi:M\longrightarrow Z_{m-1}$ is horizontal if and
only if
either of the maps $I^\pm\circ\Phi:M\longrightarrow Z_m$ are horizontal.
We
shall accomplish the inductive step by showing that any element
$\Phi$ of $\HH_d(M,Z_m)$ may be deformed into $Z^-$.

If $\Phi(M)\subseteq S^Q(Z^-)$, then $\{(\exp\,itQ)\nat\Phi\}_{0\le
t\le\infty}$ provides a continuous deformation of $\Phi$ into $Z^-$.
So it suffices to show that $\Phi$ can be deformed into $S^Q(Z^-)$.
We claim that there exists some $A\in SO_{2m+1}$ such that
$A\nat\Phi(M)\subseteq S^Q(Z^-)$, i.e. $A\Phi(z)\not\supseteq L$ for all
$z\in
M$; this will give the required deformation, as $SO_{2m+1}$ is
connected. Since $SO_{2m+1}$ acts transitively on the space $Y_m$
of all isotropic lines in $\C^{2m+1}$, it suffices to find some
isotropic line $L^\prime$ such that $\Phi(z)\not\supseteq L^\prime$ for all
$z\in M$. Let
$$
Y_m^\Phi=\{L^\prime\in Y_m\ \vert\ L^\prime\subseteq\Phi(z)\ \text{for
some}\ z\in M\}.
$$
Then our claim is that $Y_m^\Phi\ne Y_m$. Let
$X_m=\{(L^\prime,W)\in Y_m\times Z_m\ \vert\ L^\prime\subseteq W\}.$
Let $p_1,p_2$ be the projections to $Y_m, Z_m$. Then
$Y_m^\Phi=p_1(p_2^{-1}(\Phi(M)))$. We have $\dim_{\C}Y_m^\Phi
\le\dim_{\C} p_2^{-1}(\Phi(M))\le m-1+\dim_{\C}\Phi(M)$
(as the fibre of $p_2$ is $\C P^{m-1}$) $\le m$. Since $\dim_{\C}
Y_m=2m-1$, $Y_m^\Phi$ cannot be equal to $Y_m$ if $m\ge2$. This
completes the proof.
\qed\enddemo

It should be clear from these examples that a similar method applies to
those harmonic maps $\varphi:M\lra G/K$ which are of the form
$\varphi=\pi\circ\Phi$, where $\Phi$ is holomorphic and super-horizontal
with respect to a twistor fibration $\pi:G/H\lra G/K$. That is, for a height
function $h^Q:G/H\lra\R$ (where $G/H=\Ad(G)P$), we obtain deformations
$\Phi_t$ of $\Phi$ such that $\Phi_\infty$ takes values (generically) in a
critical manifold $C(T_Q)/C(T_Q)_X=\Ad\,C(T_Q)X$ of $h^Q$. To obtain a
continuous deformation, one must ensure that the image of $\Phi$ lies
entirely in the stable manifold of this critical manifold.

Without loss of generality we may assume that $X=P$. A calculation similar
to that of Lemma 3.7 then shows that the bundle $C(T_Q)/C(T_Q)_X\lra
C(T_Q)/C(T_Q)_X\cap K$ is a \lq\lq twistor sub-fibration\rq\rq\ of
$G/H\lra G/K$.
Lemma 3.7 provides an infinite dimensional version of this phenomena ;
namely, that the bundle $G/H\longrightarrow G/H$
may be regarded as a twistor sub-fibration of the fibration
$\Omega G\longrightarrow G$.
As explained in \S 3, $\blg$ can be realised as the
orbit of the point $\al=(i,0)\in i\R\ltimes \La\frak g$, under the
action of $S^1\ltimes\flg$. The theory described in the Appendix for a
finite dimensional adjoint orbit extends almost entirely to $\blg$
(cf. ~\cite{AP}, \S 8.9 of \cite{PS}, and \cite{Ko}), although there are some
new
features. For example, the inner product $\langle\ ,\ \rangle$ is not
bi-invariant with respect to the action of $S^1\ltimes\flg$. From
our point of view, the main difference is that it is not in general
possible to integrate the gradient vector field on the infinite dimensional
manifold $\blg$. In fact (Theorem 8.9.9 of \cite{PS}), every
point $\ga\in\blg$ admits a \ll downwards\rr flow line, but only points of
$\ablg$ admit \ll upwards\rr flow lines. The
asymmetrical nature of the flow reflects the fact that the
action of $\CS$ on $\ablg$ extends to an action of $\C^\ast_1=\{\la\in\CS\
\vert\ \vert\la\vert<1\}$ on $\Omega G$, but not to an action of $\CS$.

 \newpage
\heading
Appendix: Height functions on generalized flag manifolds
\endheading

Let $G$ be a compact connected Lie group. The orbit $M_P=\Ad(G)P$ of a
point
$P\in\frak g$ under the adjoint representation is called a
generalized flag manifold. It is known that the isotropy subgroup of
$P$ is the centralizer, $C(T_P)$, of that torus $T_P$ which is
the closure of the one parameter subgroup $\{\exp\,tP\}$. The
complex group $\gc$ also acts transitively on $M_P$, and the
isotropy subgroup of $P$ is a parabolic subgroup $G_P$ of $\gc$.
Thus, we have natural diffeomorphisms
$$
M_P\cong G/C(T_P)\cong\gc/G_P.
$$
We denote the natural action of $\gc$ on $M_P$ by $(g,X)\longmapsto
g\nat X$. (If $g\in G$, then $g\nat X=\Ad(g)X$.)
The standard example of this is given by $G=U_n$ and
$P=i\pi_V\in\frak u_n$, where $V$ is a $k$-dimensional subspace of
$\C^n$ and $\pi_V$ denotes orthogonal projection from $\C^n$ to $V$
with respect to the Hermitian inner product of $\C^n$. Then
$$
M_P\cong U_n/U_k\times U_{n-k}\cong \glnc/G_P,
$$
where $G_P=\{A\in\glnc\ \vert\ AV\subseteq V\}$. This can be identified
with the Grassmanian $\grkcn$, by identifying $\Ad(A)P$ with the
$k$-plane $AV$. The action of $\glnc$ on $\grkcn$ is then given by
the formula $A\nat V=AV$.
The homogeneous space $M_P$ has a natural K\"ahler structure,
which is determined by the choice of $P$ and a choice of an $\Ad(G)$-
invariant
inner product $\langle\langle\ ,\ \rangle\rangle$ on $\frak g$.

For any $Q\in\frak g$, we define the \ll
height function\rr $h^Q:M_P\longrightarrow\R$ by
$$
h^Q(X)=\langle\langle X,Q\rangle\rangle.
$$
A point $X\in M_P$ is a critical point of $h^Q$ if and only if
$[Q,X]=0$. It follows from this that the critical points of $h^Q$ form
a finite number of orbits of the group $C(T_Q)$, say
$$
N_{1}=\Ad(C(T_Q))X_1,\quad\dots\quad,N_{r}=\Ad(C(T_Q))X_r.
$$
These critical manifolds are
non-degenerate; in other words, $h^Q$ is a \ll Morse-Bott function\rrr. In
the standard example, where $M_P\cong\grkcn$, let us choose
$Q=i\pi_l$ where $\pi_l:\C^n\longrightarrow\C^l$ is orthogonal projection
onto
the span of the first $l$ standard basis vectors. A point
$\Ad(A)P=i\pi_W$ is a critical point of $h^Q$ if and only if
$[\pi_l,\pi_W]=0$, i.e. $W=W_0\oplus W_1$ where $W_0\subseteq\C^l$,
$W_1\subseteq(\C^l)^\perp$. The critical manifold $N$ containing
$W=W_0\oplus W_1$ is the set of $k$-planes $U$ such that
$U=U_0\oplus U_1$, where $U_0\subseteq\C^l$, $U_1\subseteq(\C^l)^\perp$,
and $\dim\,U_i=\dim\,W_i$ for $i=0,1$. It is the orbit of $i\pi_W$
under the group $C(T_Q)=U_l\times U_{n-l}$, and hence is a copy of
$Gr_{w_0}(\C^l)\times Gr_{w_1}(\C^{n-l})$, where
$w_i=\dim\,W_i$. The index of a critical manifold may be computed using
the Stiefel diagram of $G$. This theory is due to Bott (\cite{Bo}).

Let $\nabla h^Q$ be the gradient of $h^Q$ with respect to the K\"ahler
metric. The integral curves of $\nabla h^Q$ may be calculated explicitly,
since
$$
-\nabla h^Q=JQ^\ast
$$
where $Q^\ast$ is the vector field on $M_P$ associated to the one
parameter subgroup $\{\exp\,tQ\}$. This observation is due to
Frankel (\cite{Fr}).  It follows that the flow line of $-\nabla h^Q$ which
passes through a non-critical point $X$ is
$$
t\longmapsto (\exp\,itQ)\nat X.
$$
In the standard example, the flow line of $-\nabla h^Q$ passing
through a non-critical point $i\pi_W$ is given by
$$
t\longmapsto\pi_{W_t},\ \ \ W_t=e^{-t\pi_l}W
$$
where $e^{-t\pi_l}$ is the $n\times n$ diagonal matrix with diagonal
terms $e^{-t},\dots,e^{-t}$ ($l$ times) $1,\dots,1$ ($n-l$ times).

The stable (or unstable) manifold $S^Q(X)$ (or $U^Q(X)$) of a critical
point $X$ is by definition the union of the flow lines of $-\nabla h^Q$
which converge to $X$ as $t\to \infty$ (or as $t\to-\infty$). The
stable manifold of the critical manifold $N$ is defined by
$S^Q(N)=\bigcup_{Y\in N}S^Q(Y)$, with a similar definition of
the unstable manifold $U^Q(N)$. Using the above description of the
flow lines, it can be shown that
$$
S^Q(N)=(G_Q)\nat X
$$
i.e. the orbit of $X$ under the (complex) group $G_Q$. Similarly,
$$
U^Q(N)=(G^{opp}_Q)\nat X
$$
where $G^{opp}_Q$ is the \ll opposite\rr parabolic subgroup to
$G_Q$. In the standard example, the stable manifold of the critical
manifold $N$ is the set of $k$-planes $U$ such that
$\dim\,U\cap\C^l=w_0$.  This is the orbit of $W$ under the group
$G_Q=\{A\in\glnc\ \vert\ A(\C^l)\subseteq\C^l\}$. The unstable manifold
is the set of $k$-planes $U$ such that
$\dim\,U\cap(\C^l)^\perp=w_1$, i.e. the orbit of $W$ under the group
$G^{opp}_Q=\{A\in\glnc\ \vert\ A(\C^l)^\perp\subseteq(\C^l)^\perp\}$.

\enddocument
\end